\documentclass[aps,twocolumn,groupedaddress,10pt]{revtex4}
\usepackage{graphicx}
\usepackage[english]{babel}

\begin{document}

\title{ELM processes and properties in 2T 2MA ITER-like wall
  JET plasmas}

\author{A. J. W\fontshape{sc}\selectfont{ebster}$^{1,2}$}
\author{S. J. W\fontshape{sc}\selectfont{ebster}$^{1,2}$}
\author{JET EFDA Contributors\footnote{See the Appendix of
    F. Romanelli et al., Proceedings of the 24th IAEA Fusion Energy
    Conference 2012, San Diego, US.}}
\affiliation{$^1$JET-EFDA, Culham Science Centre, Abingdon, OX14 3DB, UK}
\affiliation{$^2$ EURATOM/CCFE Fusion Association,
  Culham Science 
  Centre, Abingdon, OX14 3DB, UK}

\date{\today}
\email[]{anthony.webster@ccfe.ac.uk}

\begin{abstract}
During July 2012, 150 almost identical H-mode
plasmas were consecutively created in the Joint European Torus (JET),
providing a combined total of approximately 8 minutes of steady-state
plasma with 15,000 Edge Localised Modes (ELMs).  
In principle, each of those 15,000 ELMs are statistically equivalent. 
Here the changes in edge density and plasma energy associated with
those ELMs are explored, using the spikes in Beryllium II (527 nm)
radiation as an indicator for the onset of an ELM. 
Clearly different timescales are observed during the ELM process. 
Edge temperature falls over a 2ms timescale, edge density and pressure
fall over a 5ms timescale, and there is an additional 10ms timescale
that is consistent with a resistive relaxation of the plasma's edge. 
The statistical properties of the energy and density losses due to
the ELMs are explored. 
For these plasmas the ELM energy ($\delta E$) is found to be
approximately independent of the time between ELMs, despite the 
{\sl average} ELM energy ($\langle E \rangle$) and {\sl average} ELM
frequency ($f$) being consistent with 
the scaling of $\langle \delta E \rangle \propto 1/f$.  
Instead, beyond the first 0.02 seconds of waiting time between ELMs,
the energy losses due to individual ELMs are found to be statistically the
same.  
Surprisingly no correlation is found between the energies of
consecutive  ELMs either. 
A weak link is found between the density drop and the ELM waiting 
time. 
Consequences of these results for ELM control and modelling are
discussed. 
\end{abstract}


\maketitle

\section{Introduction}

Edge Localised Modes (ELMs) are instabilities that occur at the edge 
of tokamak plasmas \cite{Wesson}.  
They are thought to be triggered by an ideal Magnetohydrodynamic (MHD)
instability of the plasma's edge \cite{EPED1,WebsterNF}, and are 
presently found in nearly all high confinement tokamak
plasmas \cite{Keilhacker84,Zohm, Kamiya}.  
Large ELMs such as those that are predicted to occur in ITER 
\cite{Aymar,Lipschultz,Loarte}, 
will need to be reduced in size or avoided entirely if plasma-facing
components are to have a reasonable lifetime. 
One way to reduce ELM size is by ``pacing'' the ELMs at
higher frequencies than their natural rate of occurrence 
\cite{Liang,LangPacing},  
because 
they are expected to occur with a lower energy due to the empirically
observed relationship between ELM energy ($\delta E$) and ELM
frequency ($f$) of $\delta E \propto 1/f$ \cite{Hermann}. 
The ELM frequency is usually reported 
as an average over all ELMs in a given pulse, and is identical to
one divided by the 
average waiting time between the ELMs. 
In contrast to the relationships between the average ELM energy and
average ELM frequency, the relationship between an individual ELM's energy
and its individual ``frequency'' 
(often defined as one divided by its waiting
time since the previous ELM), is rarely reported.
It is this topic that is considered here.

Since 2011 the JET tokamak has been operating with its previously
Carbon plasma-facing components replaced with the metal ITER-like wall
\cite{Rudi}. 
This has led to differences in plasma confinement and ELM properties,
as discussed for example in \cite{Rudi, Maddison} and references
therein. 
This paper focuses its attention on a set of 150 JET plasmas produced
over a two week period in July 2012, 120 of which were nearly
identical, providing $\sim$10,000 statistically equivalent ELMs. 
Such high quality statistical information on ELM properties has never 
previously been available. 
The pulses are 2 Tesla 2 Mega Amp H-mode plasmas with approximately
12MW of NBI heating, a fuelling rate of 1.4$\times 10^{22}$ Ds$^{-1}$,
Z$_{eff}$=1.2, and
a triangularity of $\delta$=0.2, see \cite{NewRef2} for further
details, including a large selection of time traces. 
The plasmas each have
approximately 6 seconds of steady H-mode, 2.3 seconds of which between
11.5 and 13.8 seconds is exceptionally steady and is what we consider
here and in previous work \cite{EPSResonances, Resonances}.

\begin{figure*}[htbp!]
\begin{center}
\includegraphics[width=15cm]{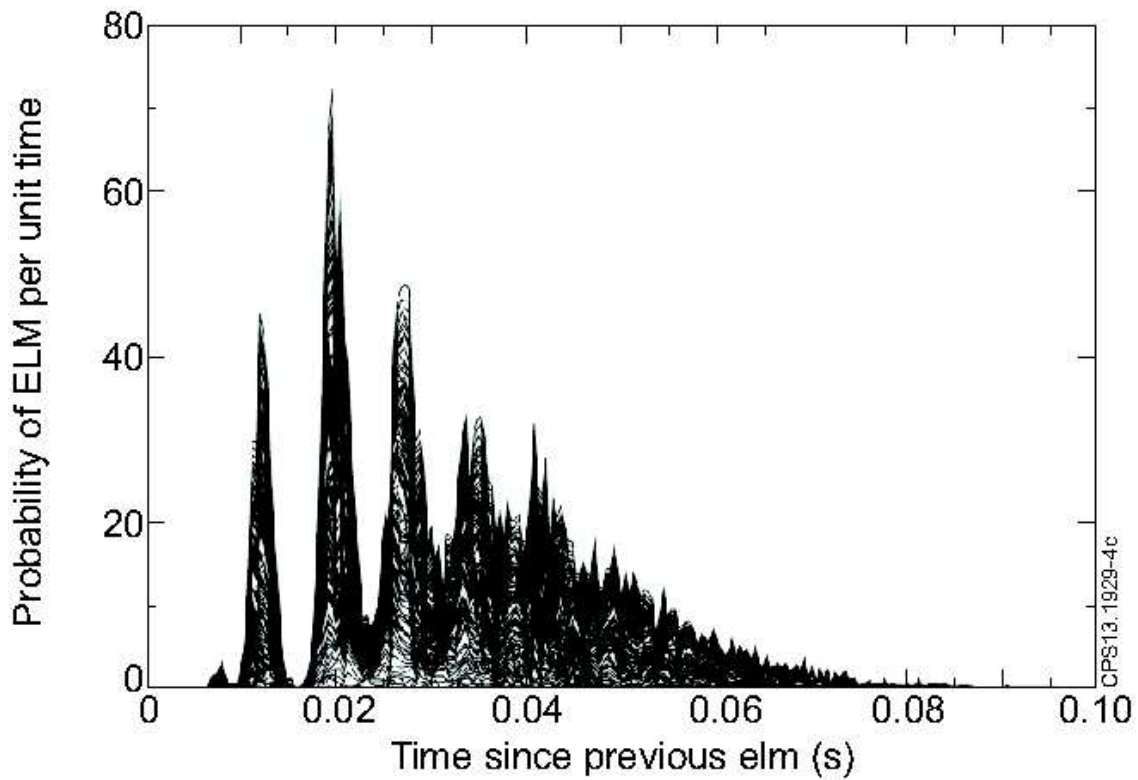}
\end{center}
\caption{ \label{Tree}
The probability density function (pdf) for the waiting time between
ELM events, determined from the ELM waiting time data from 120
equivalent pulses (see text for details). 
Each line corresponds to data from an individual pulse.  
Reproduced from Ref. \cite{Resonances}.
} 
\vspace{0.5cm}
\end{figure*}

The large quantities of extremely high quality steady-state data that
these plasmas provide, allows statistical methods to 
observe details that could not otherwise be seen, such as an
unexpected series of maxima and minima in the probability density
function (pdf) for the waiting times between ELMs that was created
from the experimental data (see figure \ref{Tree}, reproduced from 
Ref. \cite{Resonances}). 
The series of maxima and minima in figure \ref{Tree} are not due to
different ELM frequencies in different pulses, but arise from a
sequence of statistically almost independent ELM events, whose
resulting probability distribution is in figure \ref{Tree}.  
The cause of this phenomenon is not fully understood, and is presently
under investigation, preliminary results are in
Refs. \cite{EPSResonances} and \cite{Resonances}. 
A question that motivated this paper was whether the maxima and minima
in figure \ref{Tree} had a similar distribution of "quantised" ELM
energies.  
The answer we will find is no, but the excellent statistical
information has led to a number of other surprising results that we
will report here.   
It is worth noting that it requires of order 250-500 ELMs to clearly
observe the 4-5 maxima and minima of figure 1. 
This 
typically will require pulses to
be repeated 4-5 times, and considerably more if we are to ensure that
the statistical noise is kept small. 
It also requires pulses that are
extremely steady. 
Such large quantities of high quality data are not
generally available, and as a result, it is not possible at present to
be certain about how common the phenomenon is, or whether it is only
present in ITER-like wall plasmas. 
We will find no evidence that the phenomenon is affecting the ELM
energies at all, it seems solely to affect the times at which ELMs are
triggered. 
Therefore the phenomenon is not discussed further here, but seems likely to
be important for understanding how ELMs are triggered. 
The outline of the paper is as follows. 
In Section \ref{define} we describe how we determine and define 
individual ELM sizes. 
In Section \ref{ELMstats} we describe the statistical properties of
the ELMs.
Section \ref{dtp} considers the average evolution of the edge
temperature and pressure, 
 and in Section \ref{Concs} we discuss the results and
propose our conclusions. 

\section{Defining the energy and density drop due to an
  ELM}\label{define} 

\begin{figure*}[htbp!]
\begin{center}
\includegraphics[width=15cm]{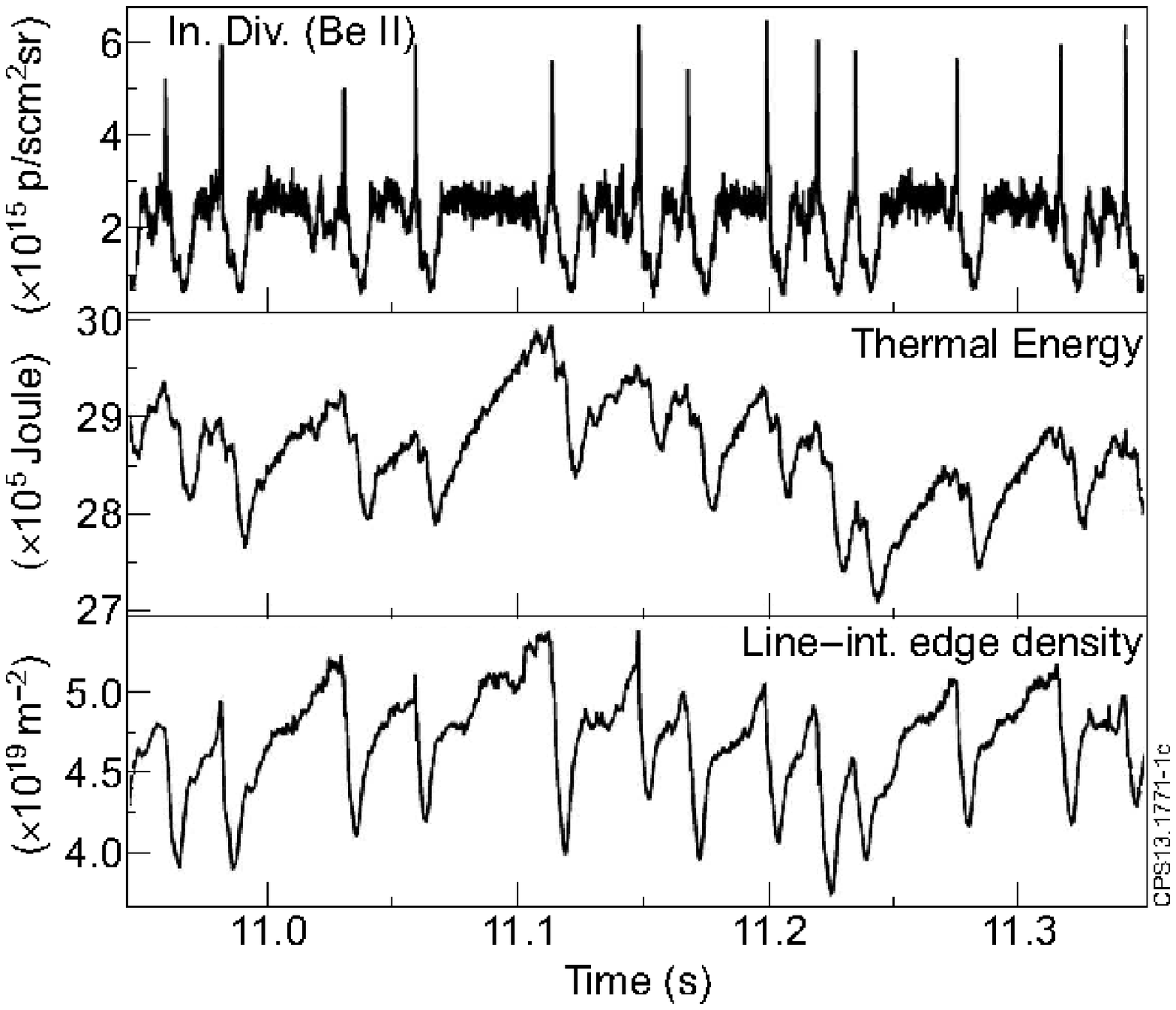}
\end{center}
\caption{ \label{Signals}
From top to bottom: i) The Be II (527nm) signal measured at the inner
divertor, that is used to identify ELMs from the sharp spikes in
radiation. ii) The estimated thermal energy stored in
the plasma, which is calculated by EFIT using magnetic measurements to
reconstruct the MHD equilibrium and infer the plasma's pressure. The
thermal energy is $3/2$ times the volume integral of the plasma pressure,
sometimes referred to as the plasma's ``kinetic'' energy. iii) The  
line-integrated plasma number density at the 
plasma's edge. 
For each ELM there is a sharp spike in Be II emission, shortly
followed by a drop in density to a minimum at around 0.005s after the
ELM started, and a drop in the plasma's thermal energy to a minimum
some time around 0.01s after the ELM started. 
} 
\vspace{0.5cm}
\end{figure*}

The main purpose of this paper is to explore the relationship between
the losses of plasma energy or density due to ELMs, and the waiting
times between the ELMs. 
The signals that are used are the line integrated edge plasma density,
which is a direct line-integrated measure of the density at the plasma's
edge, and the plasma's thermal energy as inferred from a collection
of magnetic diagnostics using EFIT \cite{EFIT,EFIT2}, 
both of which are checked and compared with independent Thompson
scattering measurements. 
The Beryllium II (527nm) radiation that is measured at the inner
divertor is used to determine when ELMs occur, using 
the method described in \cite{WebsterDendyPRL}, that detects the statistically
large  spikes in  radiation that are associated with ELMs. 
For the type I ELMs in the H-mode plasmas considered here, the ELMs
are easy to identify with this method.  
All the signals just described are standard and widely used JET
signals.
An advantage of the signals chosen is that they are independent
measurements for: the line integrated density, the Be II light emissions,
and the thermal energy losses inferred from magnetic measurements; and the
former two measurements are direct measurements requiring minimal
reconstruction from diagnostic data. 
Following an ELM there is a small radial plasma motion by 7-8mm, that
in principle can affect measurements. 
Appendix \ref{ROG} confirms that for the plasmas studied here at
least, this can be neglected in comparison to the much larger changes in
the post-ELM measurements. 
Next we will firstly discuss the measured changes to the line
integrated plasma density, we will find that similar remarks 
apply to changes in the plasma's energy.

\begin{figure*}[htbp!]
\begin{center}
\includegraphics[width=12cm]{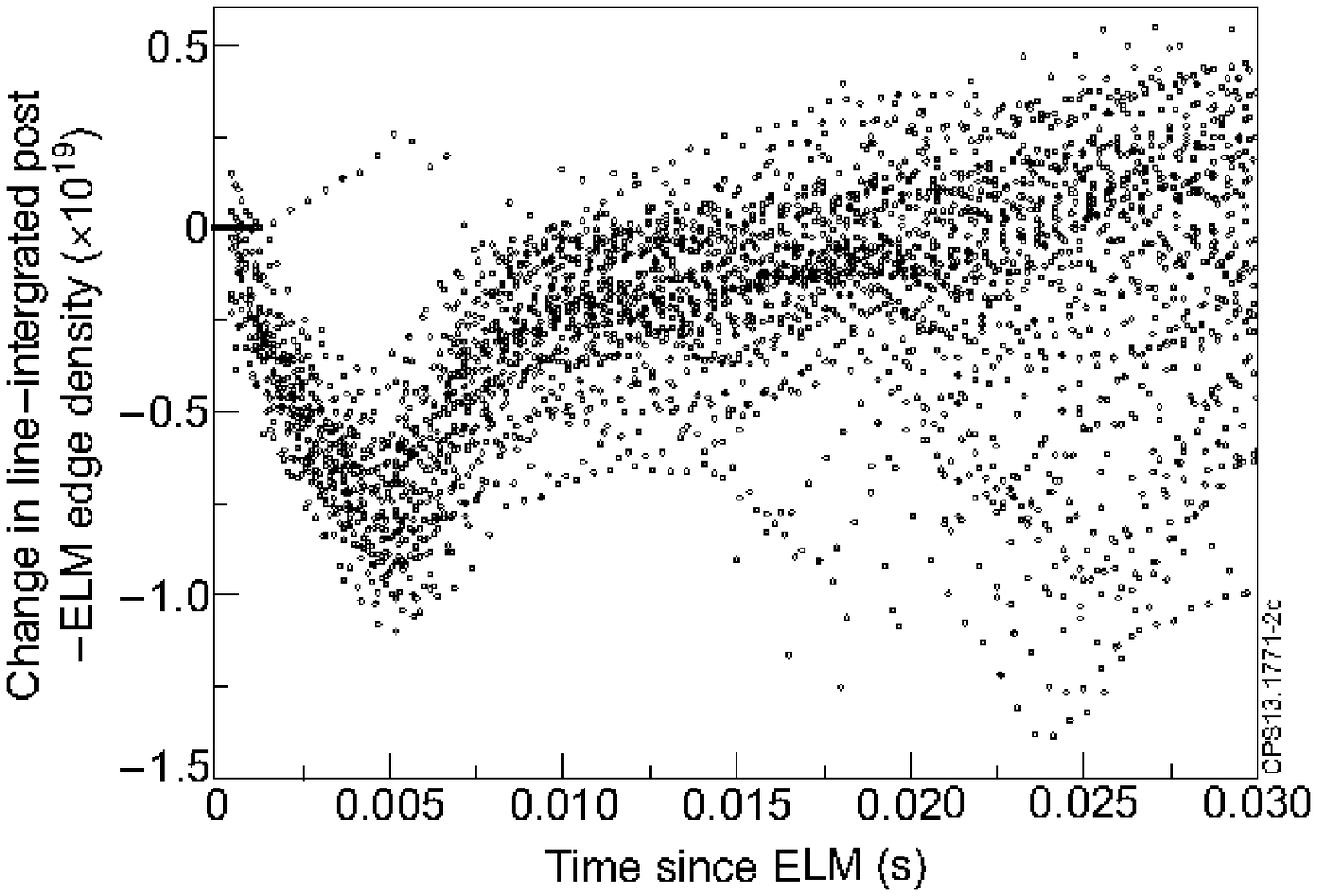}
\end{center}
\caption{ \label{PostELMdensity}
The fall in line-integrated edge density with time since each ELM is 
plotted for a typical pulse in the set (pulse number 83790).
There is a clearly visible minima at around 0.005 seconds. 
Beyond about 0.012 seconds there are a small number of additional
drops in density due to ELMs that occur within the 0.03 second time
interval that is plotted.  
} 
\vspace{1.5cm}
\end{figure*}

\begin{figure*}[htbp!]
\begin{center}
\includegraphics[width=12cm]{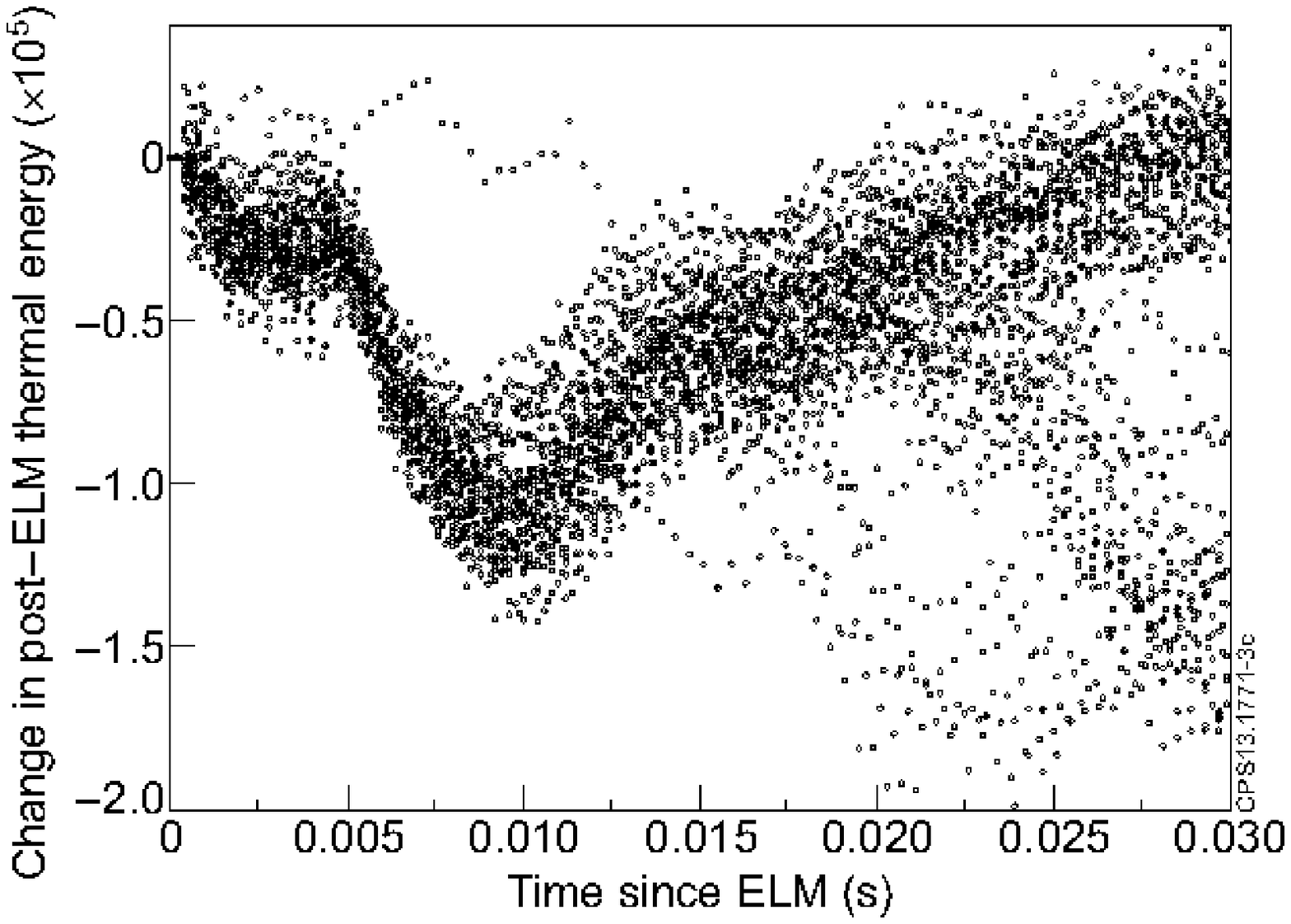}
\end{center}
\caption{ \label{PostELMEnergy}
The fall in the plasma's thermal energy with time since each ELM is 
plotted for a typical pulse in the set (pulse number 83790).
There are two clearly visible minima, one between 0.002 and 0.005
seconds, and another at around 0.01 seconds. 
Beyond about 0.012 seconds there are a small number of additional
drops in energy due to ELMs that occur within the 0.03 second time
interval that is plotted.  
} 
\end{figure*}

\begin{figure*}[htbp!]
\begin{center}
\includegraphics[width=12cm]{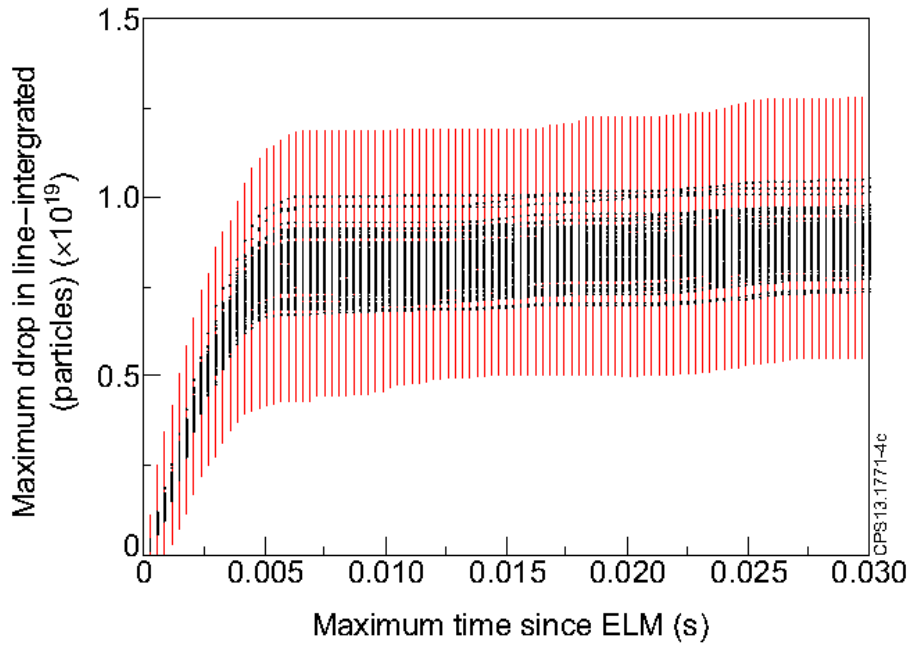}
\end{center}
\caption{ \label{dn-tm}
The maximum drop in line integrated plasma density ($\delta n$) following
an ELM (vertical axis), is plotted against the (maximum) time $t_m$
since the ELM (horizontal axis), over which the maximum drop is
calculated.  
For each plasma $\delta n$ is averaged over all the ELMs in a given
pulse (plotted points), and its standard deviation 
calculated (vertical lines). 
This is repeated for each maximum time $t_m$ since the ELM, and for
each plasma pulse.  
There is a comparatively small scatter of about 15-20\% between the
average value's of $\delta n$ for the 120 different pulses, confirming
that the pulses are quite similar.  
Consequently if $t_m$ is taken to be greater than about 0.005 seconds
then there is a well defined $\delta n$ that is independent of $t_m$. 
The timescale of 0.005 seconds is much less than the time between ELMs.   
}
\end{figure*}

Following an ELM, the line integrated plasma density falls, then
recovers again (see figure \ref{Signals}). 
The losses associated with the ELM have 
a duration  of order 0.005 seconds, that
combined with fluctuations in the signal can make it difficult to define
the density loss due to the ELM. 
For example, figure \ref{PostELMdensity} shows the fall in  edge
density with time since an ELM for ELMs in the typical pulse 83790. 
There is clearly a minimum in the line integrated signal at around
0.005 seconds,  or in equivalent words, there is a maximum drop in
line integrated edge density at around 0.005 seconds. 
The exact time
and magnitude of the minimum is not always the same. 
Here we define the density drop due to an ELM ($\delta n$) as the
maximum observed 
drop in the line-integrated density within a small time interval
$t_m$ after an ELM (see figure \ref{dn-tm}).
Note that figures \ref{Signals}, \ref{PostELMdensity}, and
\ref{PostELMEnergy} discuss time traces in which there 
are minima in line-integrated density or thermal energy after an ELM,
whereas from figure \ref{dn-tm} onwards we consider the maximum energy
and density lost after an ELM, which is a positive quantity.
Figure \ref{dn-tm} shows that if $\delta n$ is defined in this way
then provided $t_m$ is greater than about 0.005
seconds, which is much less than the 0.012 second waiting time to the
most frequent ELMs \cite{EPSResonances,Resonances}, then $\delta n$ is
independent of $t_m$.  
Consequently provided $t_m$ is greater than 0.005 seconds, then
$\delta n$ is independent of $t_m$ and is well
defined.  
For plots involving drops in edge density we use $t_m =$ 0.01 seconds,
and for plots involving drops in energy we will specify whether we are
discussing results with $t_m=$ 0.01 or $t_m=$0.005 seconds.

\begin{figure*}[htbp!]
\begin{center}
\includegraphics[width=12cm]{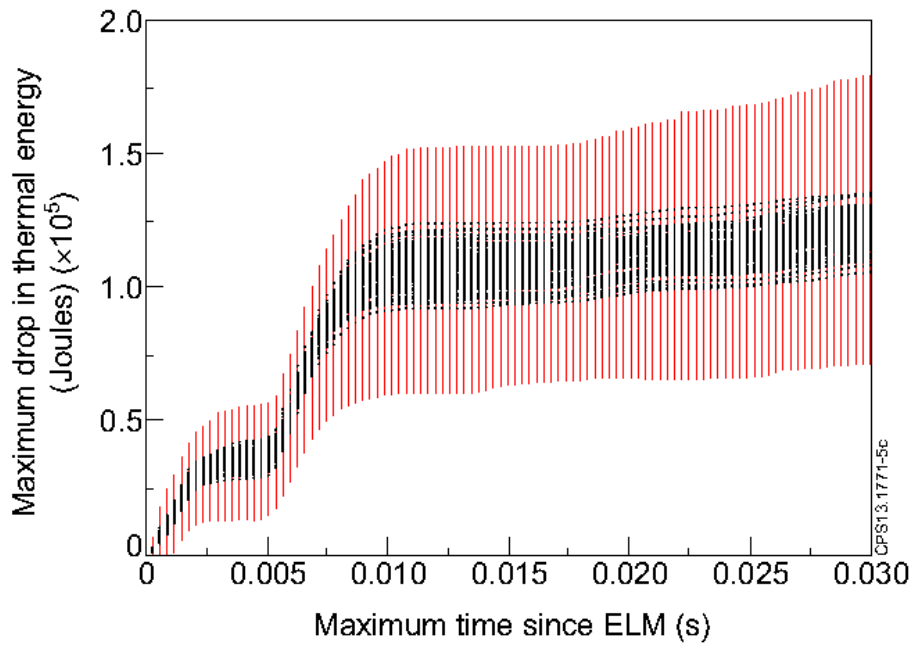}
\end{center}
\caption{ \label{dE-tm}
For  $t_m$ from 0 to 0.03 seconds and for all 120 pulses, the minimum
thermal energy within time $t_m$ after an ELM is averaged over all 
ELMs in a pulse (plotted points), and the standard deviation about
this average is calculated also (vertical lines). 
The 120 sets of averages (plotted points) and standard deviations (red
lines) have been plotted
over one another so as to present them on a single graph.  
Two time scales are evident. The first at 0.005 seconds is the same as
found in figure \ref{dn-tm}. The second timescale is at 0.01 seconds,
beyond which the average value of $\delta E$ is approximately
constant,  independent of $t_m$. 
}
\end{figure*}

Similar remarks apply to the plasma's thermal energy, which is defined
as $3/2$ times the volume integral of the plasma's pressure, with the
pressure here 
obtained from an ideal MHD reconstruction of the equilibrium using
EFIT \cite{EFIT,EFIT2}.
The thermal energy is sometimes referred to as ``kinetic energy'', but
does not include the energy due to macroscopic flows in the plasma.   
The drop in thermal 
energy ($\delta E$) is defined as the minimum energy in some time
period $t_m$ immediately following an ELM. 
A difference is that 
there are now two timescales that can clearly be observed (see figures 
\ref{PostELMEnergy} and \ref{dE-tm}). 
The first minimum in energy occurs between 0.002 and 0.005 seconds,
which tends to be before 
the minima at 0.005s found in figure \ref{dn-tm}. 
However, unlike the density, there is a second minimum at around
0.01 seconds (see figure \ref{PostELMEnergy}). 
The possible causes of the different timescales  are discussed in
greater detail later. 
Beyond 0.01 seconds the average of $\delta E$ is approximately
independent of $t_m$, allowing $\delta E$ to be defined as either the
minimum thermal energy in the time interval between an ELM and $t_m =
0.005$ seconds or between an ELM and $t_m$=0.01 seconds (see
figure \ref{dE-tm}). 
Both of these are less than the time of the first maxima in the ELM
waiting time distribution \cite{EPSResonances,Resonances}, that is at
approximately 0.012 seconds.  
This suggests two possible definitions for the ELM energy, as either the
maximum energy lost over the 0.005 second timescale during which
particle loss is also leading to a reduction in the edge density (see
figures \ref{dn-tm} and \ref{dE-tm}), or
as the total reduction in stored thermal energy over 0.01 seconds. 
Both will be reported and discussed here, and both can be observed in
the time traces in figure 
\ref{Signals}, with a small minimum in $\delta E$ prior to the minimum
in the density, followed by a much larger minimum in $\delta E$ on the
larger timescale of $\sim 0.01$ seconds. 

Two timescales have previously been reported in conjunction with the
edge electron temperature during the post-ELM pedestal recovery in
ITER-like wall plasmas 
\cite{Beurskens,Frassinetti}, an important difference is that here
the two timescales are observed with every ELM.  
It is possible that the two timescales relate to a similar sequence of 
processes - rapid energy losses followed by slower transport
processes.  
The timescale for the initial fall in
edge temperature reported in Refs. \cite{Beurskens,Frassinetti} is
only about 0.002 seconds, whereas the drop in edge density (figures
\ref{PostELMdensity} and \ref{dn-tm}), is over  a 0.005 second
timescale. 
Two timescales have also been reported in conjunction with infra red
(IR) images of JET's divertor during Carbon-wall JET experiments
\cite{Eich1}. 
In this latter work the two timescales arose from the shape of the ELM 
power deposition curve with respect to time, and are much shorter than
those discussed so far. 
The timescales characterise the initial rapid rise in ELM power
deposition, over a timescale time 
$\tau_{rise}\sim 0.0002-0.0005$ seconds,
and a slower $\tau_{decay}\sim 0.001-0.0025$ seconds
that characterises the subsequent fall in the power deposition.   
The work referred to above and the results here are consistent with,
and possibly extend, the proposed sequence of steps by which energy is
lost during an ELM \cite{Loarte}.  
Firstly there is a rapid rise in heat flux that for Carbon-wall
plasmas was found over a timescale of order
0.2-0.5 milliseconds \cite{Eich1}, with heat being lost predominately
by electrons. 
In ITER-like wall plasmas, after of order 1-2 milliseconds the edge 
temperature is found to fall to a minimum 
\cite{Beurskens,Frassinetti}, something we find here also in Section
\ref{dtp}.  
This process of energy loss is referred to as ``conduction''
\cite{Loarte}. 
Next, for the plasmas described here at least, there is a loss of
ions that is completed within a timescale of order 5 milliseconds (figure
\ref{dn-tm}), in a process referred to as ``convection''
\cite{Loarte}. 
Finally we find an additional timescale of order 10 milliseconds after an
ELM (figure \ref{dE-tm}), during which EFIT \cite{EFIT,EFIT2} suggests
that the thermal plasma energy relaxes to a minimum, before starting
to rise again. 
As discussed in Section \ref{dtp}, EFIT's reconstructed measurements are
consistent with direct Thompson scattering measurements over the 0-5
millisecond time period, but disagree between 5-10 milliseconds when
EFIT suggests that the thermal energy continues to fall. 
Appendix \ref{relax} explores the timescales associated with a
resistive  relaxation of the pedestal at the plasma's edge
\cite{Frassinetti2}, and finds a timescale of 8 milliseconds, 
very similar to the 10 millisecond timescale observed in figures
\ref{PostELMEnergy} and \ref{dE-tm}. 
Consequently it is possible that a resistive mechanism is allowing the
plasma to relax to a new post-ELM equilibrium, and
that one or more non-ideal affects are  
making EFIT's ideal-MHD equilibrium reconstruction unreliable over
this longer time period. 
Similarly, resistive effects can only become important over timescales
approaching 8 milliseconds, which may explain why EFIT's calculated
pressure agrees with that measured by Thompson scattering over the
shorter 0-5 millisecond timescale (see Section \ref{dtp}). 
It is interesting to note that 8 milliseconds is the approximate time
between the maxima and minima observed in the ELM waiting time pdf in
figure \ref{Tree}
\cite{EPSResonances,Resonances}, that will be observed later in the
time periods between the clusters of 
ELMs in figures \ref{dE-dt}, \ref{dE-dt-0.005s}, and
\ref{dn-dt}. 
We do not know whether this is a coincidence or not.


\section{Statistical properties of ELMs}\label{ELMstats}

\begin{figure*}[htbp!]
\begin{center}
\includegraphics[width=12cm]{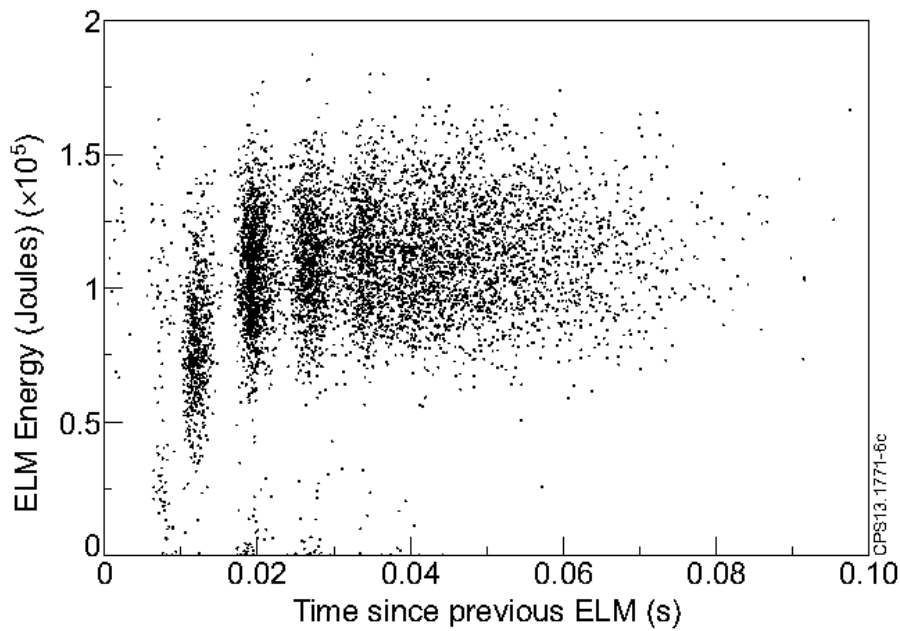}
\end{center}
\caption{ \label{dE-dt}
The drop in the plasma's thermal energy is plotted against waiting
time since the previous ELM, with $\delta E$ calculated using
$t_m =$ 0.01 seconds. 
The vertical clustering is due to the unusual ELM waiting time pdf
described in references \cite{EPSResonances,Resonances}, and shown in
figure \ref{Tree}. 
The total stored thermal energy was of order (2.8)$\times$10$^6$
Joules, so the drop in thermal energy is of order 0.04\% of the
plasma's total thermal energy.  
Beyond 0.02 seconds the drops in energy are approximately independent of
the waiting time between the ELMs. 
}
\end{figure*}

\begin{figure*}[htbp!]
\begin{center}
\vspace{1.0cm}
\includegraphics[width=12cm]{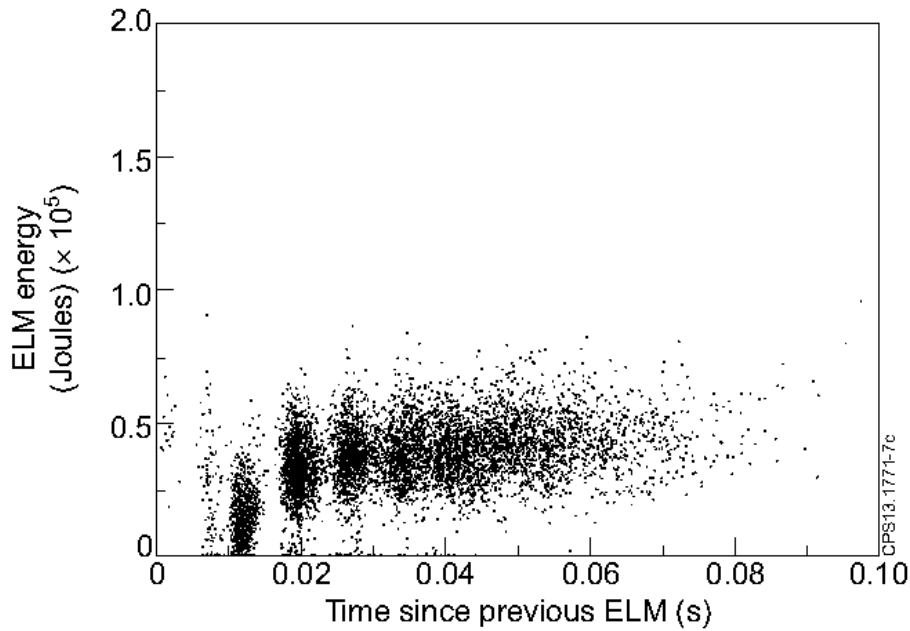}
\end{center}
\caption{ \label{dE-dt-0.005s}
The drop in the plasma's thermal energy is plotted against waiting
time since the previous ELM, as in figure \ref{dE-dt}. Here however,
$\delta E$ has been calculated using $t_m =$ 0.005 seconds, the time
of the first plateau in $\delta E$ versus $t_m$ in figure \ref{dE-tm},
and the time beyond which the drop in edge density has ended. 
}
\end{figure*}

\begin{figure*}[htbp!]
\begin{center}
\includegraphics[width=12cm]{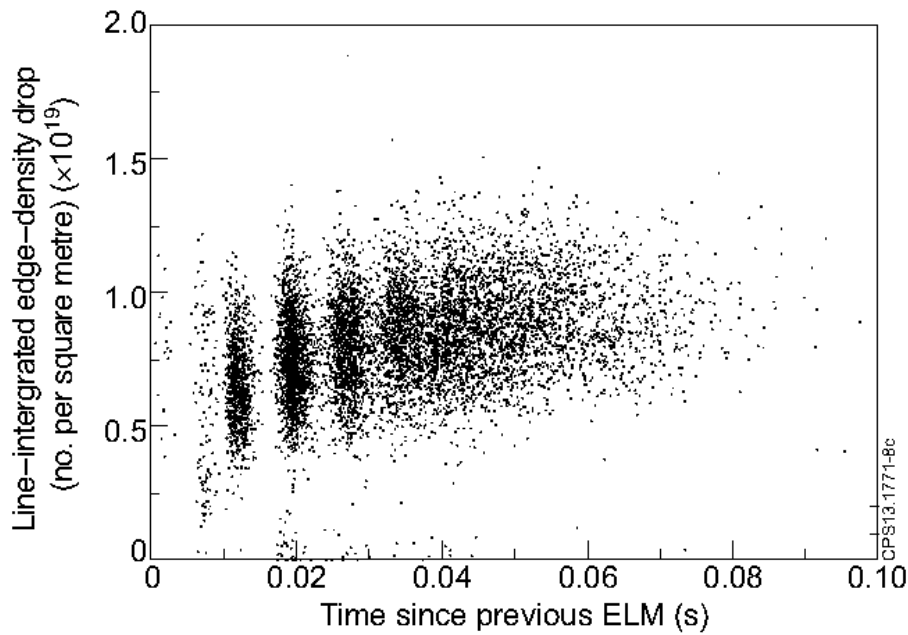}
\end{center}
\caption{ \label{dn-dt}
The drop in line integrated edge density is plotted against waiting time
since previous ELM. 
Similarly to the plot of energy against waiting time, the
vertical clustering is due to the unusual ELM waiting time pdf of the
ELMs in these pulses, as 
described in references \cite{EPSResonances,Resonances}. 
The line-integrated edge density was of order (4.5).10$^{19}$, suggesting
that roughly 20\% of the edge density is lost per ELM. 
Beyond about 0.005 seconds the minimum observed drop
in density is independent of $t_m$. 
Beyond 0.02
seconds the drop in edge density due to an ELM is only very weakly
dependent on the waiting time between ELMs.
} 
\end{figure*}

Next we look at how these measures of the density and energy losses
associated with the ELMs are influenced by the waiting times between
the ELMs (see figures \ref{dE-dt}, \ref{dE-dt-0.005s}, and \ref{dn-dt}). 
The most obvious characteristic of both figures is the vertical
clustering of ELM times. 
This is due to the waiting-time probability density function (pdf) in
figure \ref{Tree}, which is discussed in detail in 
Refs. \cite{EPSResonances,Resonances}, and shows a series of maxima
and zeros at approximately 0.08 second intervals starting from the
first maxima at 0.012 seconds and continuing until 0.04 seconds when
the distribution becomes comparatively smooth. 
The pdf was unexpected, and contrasts with large sets of ELM
waiting time pdfs that  have only  a single maxima
\cite{WebsterDendyPRL}.  
The cause of the unexpected form of pdf is unknown, and presently
under investigation. 
The next striking characteristic of figures \ref{dE-dt} and 
\ref{dn-dt}, that is particularly noticeable for the ELM energies, is
that beyond a waiting time of about 0.02 seconds the ELM energies are
similar and independent of the waiting time between the ELMs. 
In other words, the distribution of ELM energies that occur
after a waiting time of 0.02 seconds is almost identical to those of
ELMs with waiting times of 0.05 seconds or more. 
This is clearly different to the usual relationship of ELM energy
being inversely proportional to ELM frequency \cite{Hermann}, that
would lead to the  
ELM energy being linearly proportional to the ELM waiting time. 
It is also despite a continual gradual increase in edge density that
is suggested by figures \ref{PostELMdensity} and \ref{dn-dt}. 
The first large group of ELMs are observed at 0.012 seconds, and these
have an average energy that is roughly 60\% of the ELMs in later
groups. 
Similar results have been observed during pellet-triggering
experiments. 
For the specific AUG plasma scenarios reported 
in Ref. [22], a minimum waiting time of 0.007-0.01 seconds
was required before ELMs could successfully be triggered by pellets,
and beyond roughly 0.01 seconds the triggered ELMs appear to have
statistically similar energies. 
In the JET plasmas considered here, it is not known if ELMs can be
regularly triggered with waiting times less than the 0.012 second
waiting time of ELMs in the group with the highest ELM frequency
observed in figure 1. 
Pellet pacing experiments in similar 2T 2MA JET plasmas
\cite{LangJETpellets}, found a strong increase in triggering
probability for pellets at least 0.01-0.02 seconds after an ELM. 
Due to technical limitations of the pellet launcher, it was not
possible to test whether pellets could consistently pace ELMs with
waiting times of order 0.012 seconds, but the possibility of
triggering ELMs within those timescales was demonstrated. 
Therefore presuming ELMs can be paced at this 0.012
second waiting-time frequency, then an average reduction in ELM energy by
about 40\% seems a reasonable possibility. 
However there is a large scatter about the average ELM energy
for all the ELMs, independent of their waiting time,  with
standard deviations that are about 1/4 of their 
average energy. 
Consequently some of the ELMs in the 0.012 seconds waiting-time group
have ELM sizes comparable with the larger ELM sizes in the group with
longer waiting times of 0.02 seconds or more. 
Similar remarks apply to figure \ref{dE-dt-0.005s} where $t_m =$ 0.005
seconds has been used. 
The time of $t_m =$ 0.005 seconds corresponds to the first plateau of
$\delta E$ with $t_m$ in figure \ref{dE-tm}, and is the timescale over
which the edge density is lost (see figure \ref{dn-tm}). 
The group of ELMs at 0.012 seconds are about half the energy of later
ones, which is comparatively less than for figure \ref{dE-dt}, and the
overall ELM energies for waiting times greater than about 0.02 seconds
are of order 40,000 Joules.

Figure \ref{dn-dt} shows the drop in density ($\delta n$) due to the
ELMs.  Similar remarks apply as to those for  the energy losses
(figure \ref{dE-dt}), although in this case a
weak dependence of $\delta n$ on $\delta t$ remains.

So why does the observed relation between ELM energy and ELM waiting
times disagree with published studies \cite{Hermann} that find the ELM
energy ($\delta E$) to be inversely proportional to ELM frequency
($f$), with $\delta E  \propto 1/f$ ? 
It is possible that it is due to differences in behaviour between
Carbon and ITER-like wall plasmas, this remains to be determined, but
there is a simpler statistical reason that we discuss next.
The most important observation to make is that previous
studies are usually 
plotting a pulse's average ELM energy against its average 
ELM frequency, and plotting these quantities for a variety of 
different pulse types. 
In contrast, here we are plotting the individual ELM energies against
their waiting times (that can be regarded as defining 1/f for any
given ELM), and doing this for these almost identical 2T, 2MA, pulses.

\begin{figure*}[htbp!]
\begin{center}
\includegraphics[width=12cm]{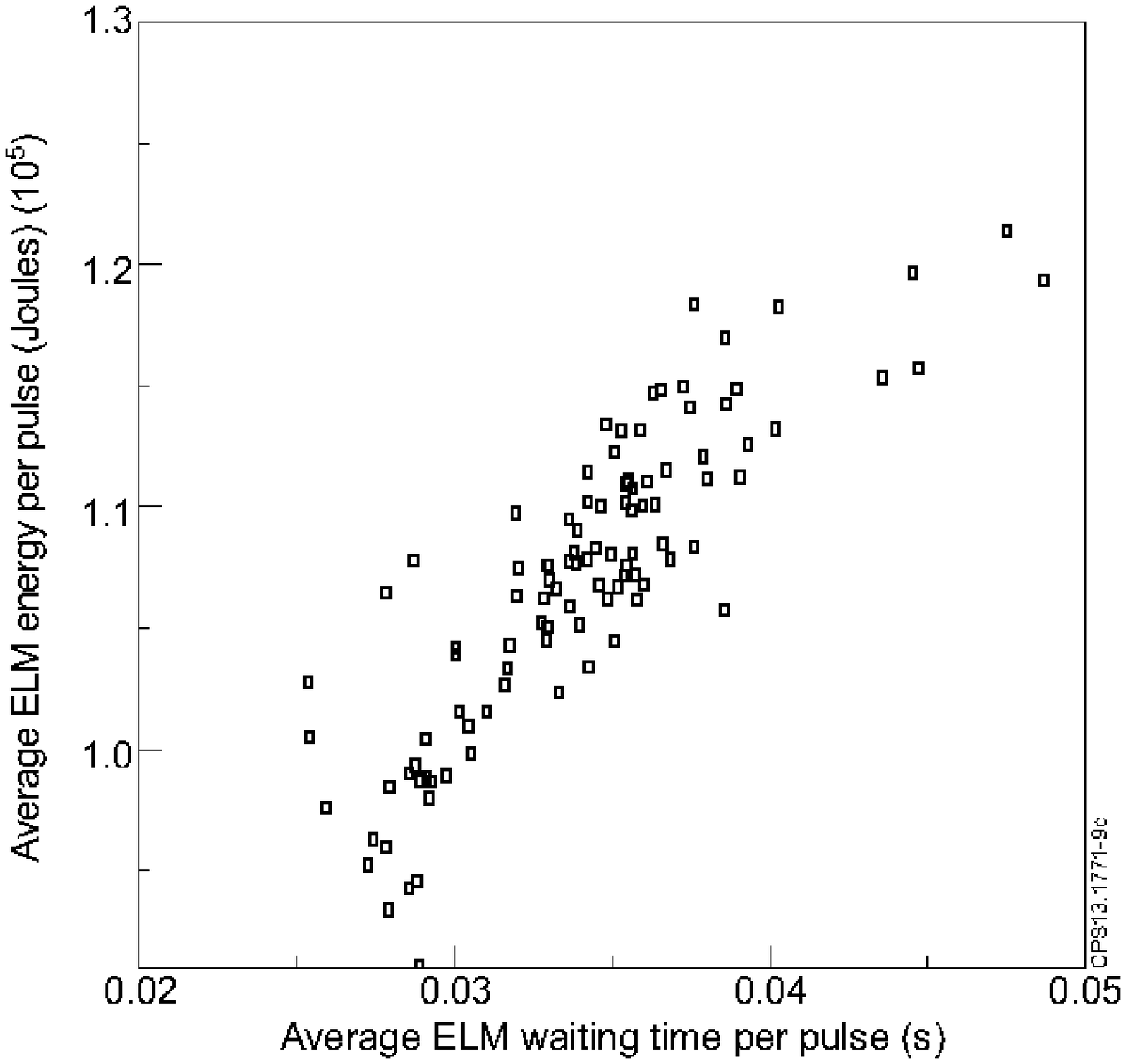}
\end{center}
\caption{ \label{AvE}
For each of the 120 pulses, the average of the thermal energy lost per ELM is
calculated using the minimum drop in energy within 0.01
seconds of the start of each ELM, and is 
plotted against the average ELM waiting time for that pulse. The
scatter is similar to what would be expected from the central
limit theorem and the roughly 50 ELMs per pulse, indicating that the
pulses are approximately statistically equivalent. The linear
relationship observed between $\langle \delta E \rangle$ and $\langle
\delta t \rangle$ is as expected if $\langle \delta E \rangle \propto
1/f$, but for the small range of $\langle \delta t \rangle$ here it is also
what would be expected from a simple Taylor expansion of $\langle
\delta E \rangle (\delta t)$, and could in principle reflect numerous
possible functions of $\delta t$. 
} 
\end{figure*}

\begin{figure*}[htbp!]
\begin{center}
\includegraphics[width=12cm]{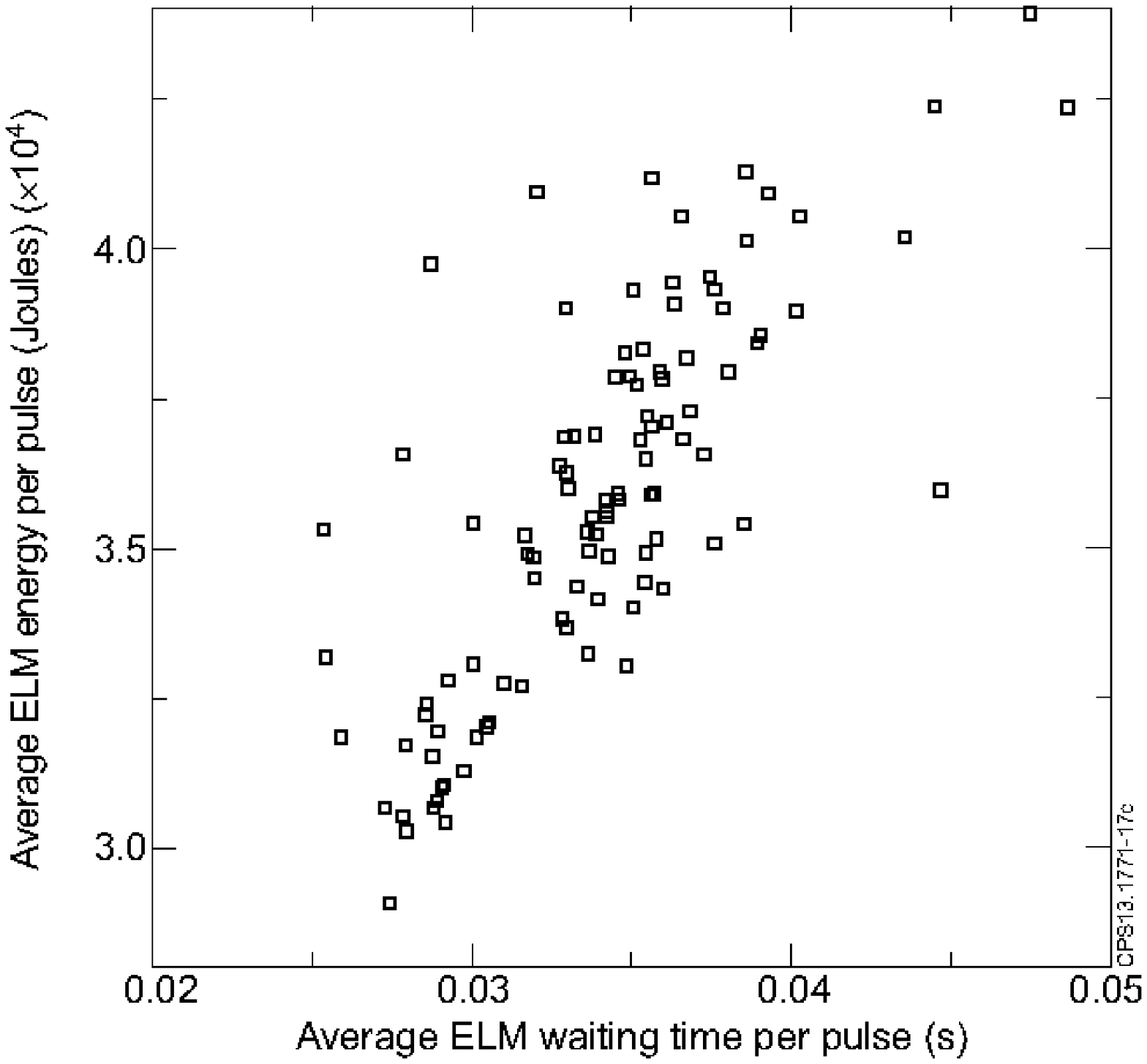}
\end{center}
\caption{ \label{AvE-0.005s}
For each of the 120 pulses, the average of the thermal energy lost per ELM is
calculated using the minimum drop in energy within 0.005
seconds of the start of each ELM, and is 
plotted against the average ELM waiting time for that pulse. The
scatter is similar to what would be expected from the central
limit theorem and the roughly 50 ELMs per pulse, indicating that the
pulses are approximately statistically equivalent.
The linear
relationship observed between $\langle \delta E \rangle$ and $\langle
\delta t \rangle$ is as expected if $\langle \delta E \rangle \propto
1/f$, but for the small range of $\langle \delta t \rangle$ here it is also
what would be expected from a simple Taylor expansion of $\langle
\delta E \rangle (\delta t)$, and could in principle reflect numerous
possible functions of $\delta t$. 
} 
\end{figure*}

If we plot $\langle \delta E \rangle$ against $\langle \delta t
\rangle$ for each of these pulses (see figures \ref{AvE} and
\ref{AvE-0.005s}), we find a simple
linear relationship that 
is consistent with $\langle \delta E \rangle \propto 1/f$, 
due to small differences in  $\langle \delta E \rangle$ and 
$\langle \delta t \rangle$ in the different pulses. 
The usual scaling between ELM energy and frequency, such as that
plotted in figure 18 of Ref. \cite{Hermann}, has 
$\langle \delta E \rangle /E
\sim 1/f \tau_E$, where $\tau_E$ is the energy confinement time of the
pulse, $\langle \delta E \rangle$ is the average thermal energy lost by
ELMs, and $E$ is the (average) thermal energy stored in the plasma.   
For the plasmas considered here, $E \sim$ 2.8$\times$ 10$^6$ J,
giving for ELM energies calculated within a 5 millisecond time period 
$\langle \delta E \rangle \sim$ (0.4 $\pm$ 0.2).10$^5$ J, 
$\langle \delta E \rangle /E \sim$ (1.3 $\pm$ 0.7).10$^{-2}$, 
$\tau_E \sim$ 0.244 $\pm$ 0.004, and $f \sim$ 31 $\pm$ 9.0, giving $f  
\tau_E \sim$ 7.6 $\pm$ 2.2, which is slightly below the scaling in 
fig. 18 of Ref. \cite{Hermann}, but the scaling is within the error
bars.  
Figure 18 of \cite{Hermann} covers roughly 2 orders of
magnitude.  
So for average ELM frequencies
at least, the results here seem consistent with the usual scaling, even if
it is not found to hold for individual ELMs within the pulses
considered here.
We note that $E/\tau_E$ is the average rate of energy loss
from the plasma, and $f\langle \delta E \rangle$ is the average rate
of energy loss by ELMs. 
Therefore if either the majority or a fixed
fraction of the energy losses are by ELMs, then the scaling of 
$E/\tau_E \sim f \langle \delta E \rangle$ (i.e. that 
$\langle \delta E \rangle/E \sim 1/f\tau_E$), is what you would
expect; only the constant  of proportionality that determines the
fraction of energy that is lost by ELMs would be expected to to change. 
However this argument only holds for the {\sl average} properties of ELMs,
not for the individual ELM energies and their individual frequencies
(the inverse of their individual waiting times), the topic that we are
most interested in here.

Exploring the statistical properties of figure \ref{AvE-0.005s} in 
more detail: 
Figures \ref{dE-tm} and \ref{dE-dt-0.005s} show that $\delta E$
has a standard deviation of order (0.2).10$^5$ Joules. 
The ELM waiting times in
figure \ref{Tree} have a standard deviation of order 0.02 seconds. 
The central limit theorem ensures that if all pulses are statistically
equivalent, then the average of n ELMs should range over an interval
whose standard deviation is a factor of $1/\sqrt{n}$ smaller in width.  
For the roughly 50 ELMs in each pulse this would lead us to expect a
range of values of $\langle \delta E \rangle$ with a standard
deviation of order (0.3).10$^4$ Joules, and values of  $\langle \delta  t
\rangle$ to have a standard deviation of order 0.003 seconds. 
This is similar to what is observed (figure \ref{AvE-0.005s}). 
Equivalent remarks apply to figure \ref{AvE}.

It could be argued that the observed linear relationship between
$\langle \delta E \rangle$ 
and $\langle \delta t \rangle$ in figure \ref{AvE-0.005s} is not
surprising.  
For the pulses here the spread of values of $\langle \delta t \rangle$
is small, with $\langle \delta t \rangle$ varying by no more than
about $\pm$ 0.01 seconds. 
Consequently it would be unsurprising if a  Taylor expansion of  
$\langle \delta E \rangle (\langle \delta t \rangle)$ were accurate
with only the linear terms in $\langle \delta t \rangle$ being kept,
consistent with 
the linear relationship observed in figure \ref{AvE-0.005s}. 
In principle the observed linear relationship could reflect numerous
possible different functions of $\langle \delta t \rangle$, not just a
linear one.  
It is possible that if the pulses were of different types with very
different values of $\langle \delta E \rangle$ and $\langle \delta t
\rangle$, then plots of $\langle \delta E \rangle$ against $\langle
\delta t \rangle$ would continue to show the linear relationship
expected if $\langle \delta E \rangle \propto 1/f$.  
However, what is clearly highlighted here is that even if the
relationship of $\langle \delta E \rangle \propto 1/f$ does hold
between different types of plasma pulses, for the plasmas studied here
at least, within a particular pulse the individual ELM energies can be
independent of their  waiting times (and the frequencies that they
define).

\begin{figure*}[htbp!]
\begin{center}
\includegraphics[width=10cm]{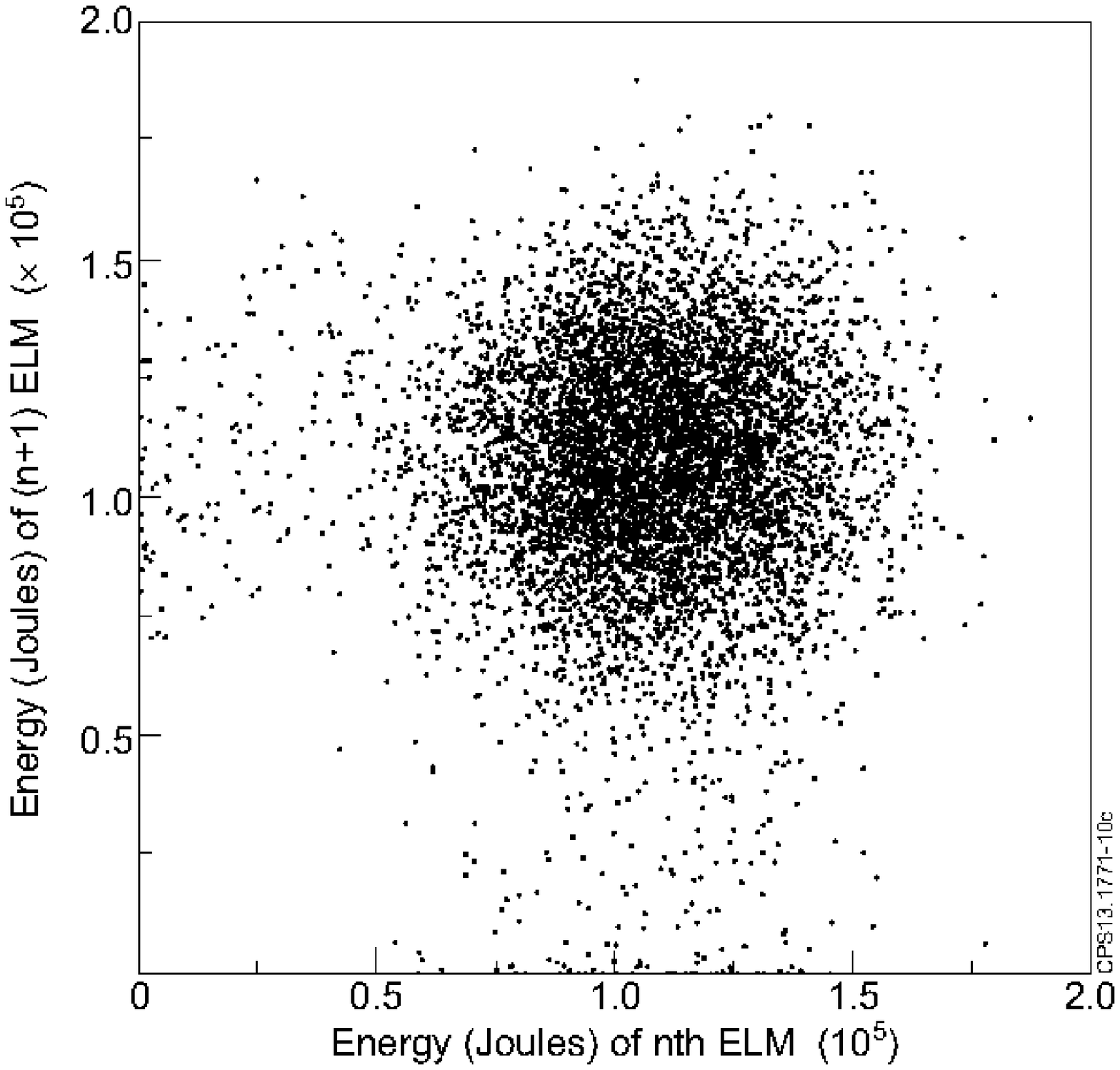}
\end{center}
\caption{ \label{Ena}
The energy of successive ELMs are plotted, with energies calculated
using $t_m =0.01$. 
Surprisingly, the clustering of subsequent ELM energies around a
single point 
indicates that the energies of subsequent ELMs are independent. 
If a large ELM were followed by a small ELM and vice versa, then we
would expect a spread of ELM energies in a perpendicular
direction to the diagonal. 
}
\end{figure*}

\begin{figure*}[htbp!]
\begin{center}
\includegraphics[width=10cm]{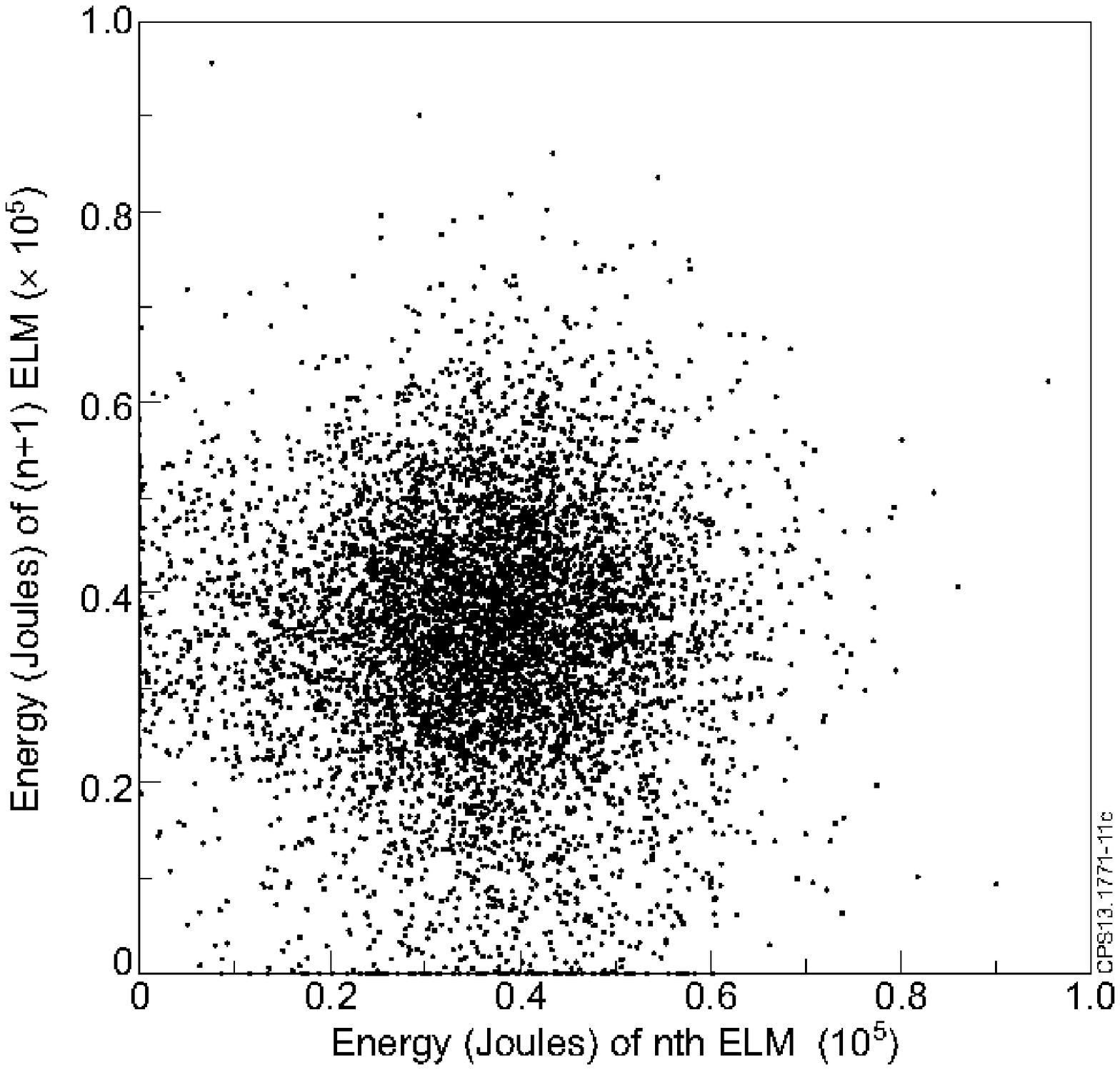}
\end{center}
\caption{ \label{Enb}
The energy of successive ELMs are plotted, with energies calculated
using $t_m =0.005$. 
Surprisingly, the clustering of subsequent ELM energies around a
single point 
indicates that the energies of subsequent ELMs are independent. 
If a large ELM were followed by a small ELM and vice versa, then we
would expect a spread of ELM energies in a perpendicular
direction to the diagonal. 
} 
\end{figure*}

A related question is whether the energies of subsequent ELMs are
related to each other, or are independent. 
For example, we might expect a large ELM to be followed by a smaller
ELM and vice versa. 
Figures \ref{Ena} and \ref{Enb} plot the energy of the nth ELM versus
the energy of the (n+1)th ELM. 
If a large ELM is followed by a smaller ELM and vice versa, then we
would expect the plotted values to cluster around a
line that is perpendicular to the diagonal. 
The symmetric clustering about an average ELM energy suggests that
the ELM energies (surprisingly) are independent.  
The same result was found for $t_m=0.01$ seconds and $t_m=0.005$
seconds, and when examining $t_{n+m}$ versus $t_n$ for $m=1$ to
$m=5$. 

\section{Edge Temperature and Pressure Evolution}\label{dtp}

\begin{figure*}[htbp!]
\begin{center}
\includegraphics[width=10cm]{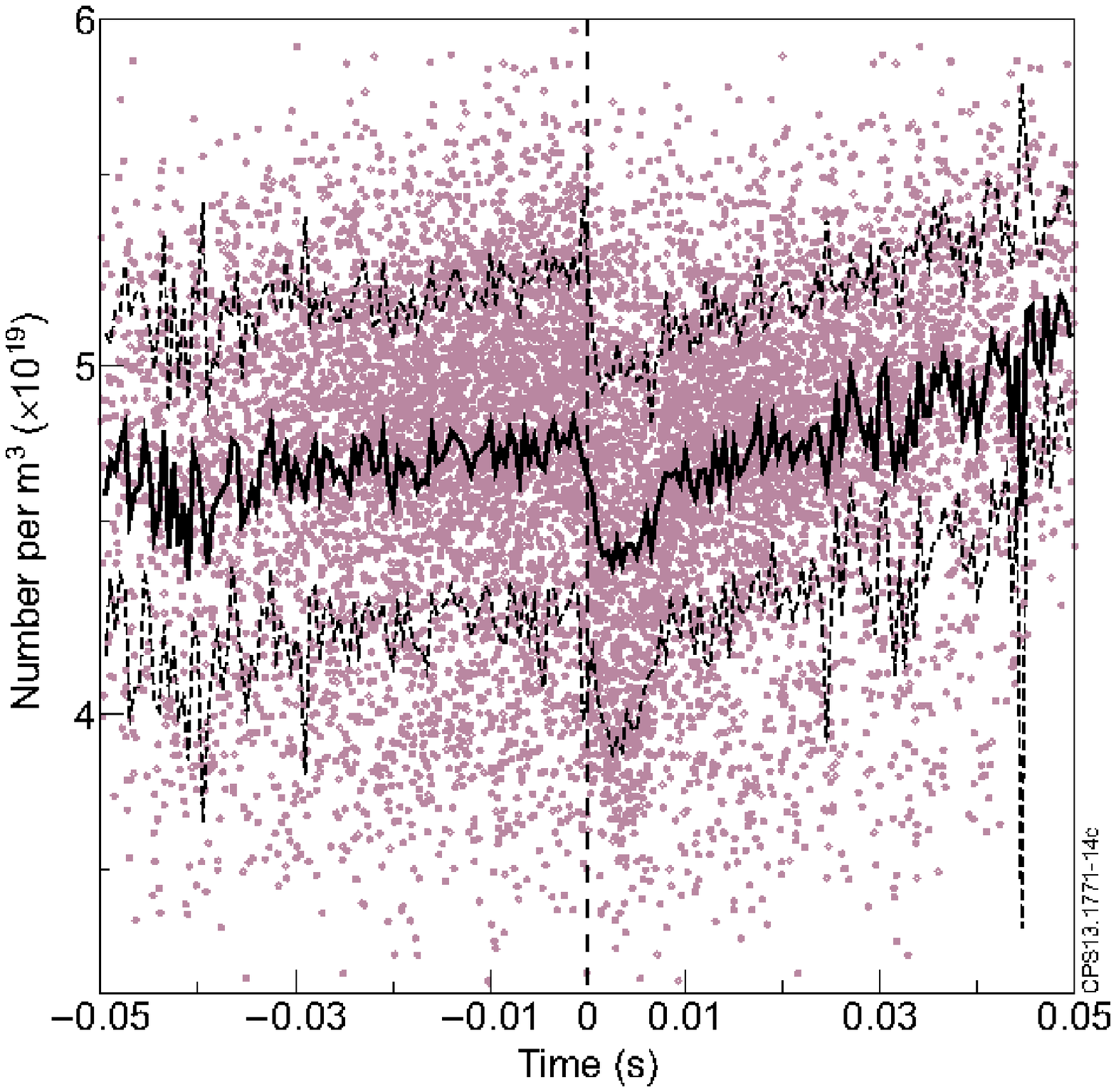}
\end{center}
\vspace{0.5cm}
\caption{ \label{dens}
The Thompson scattering measurement of particle number density,
averaged over the plasma edge region 
(between 3.74m and 3.80m along its line of sight to the magnetic
axis), from all  
ELMs and plasma pulses, synchronised to the ELM times to allow an
estimate for the density's pre- and post-ELM evolution to be made. 
Brown circles are individual measurements, the thick black line is
their average, and the dashed black lines are their average $\pm$
the standard deviation. 
The number of particles per unit volume (vertical axis) are in units
of m$^{-3}$, and the horizontal time axis is in seconds. 
Notice that there is a minimum at around 3-5 milliseconds, as was
previously observed in the line-integrated measurement (figure
\ref{PostELMdensity}). 
} 
\end{figure*}

\begin{figure*}[htbp!]
\begin{center}
\includegraphics[width=10cm]{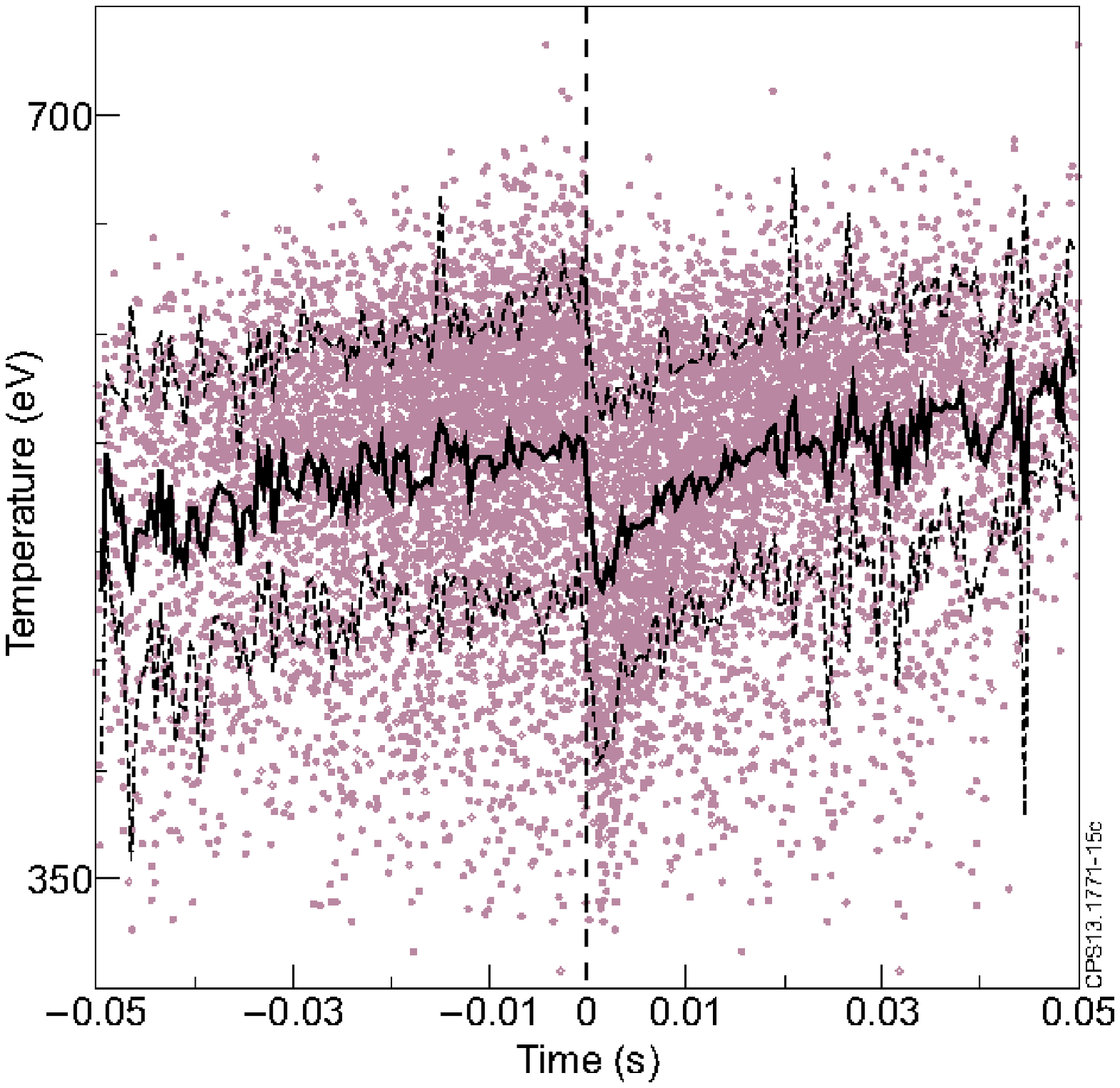}
\end{center}
\vspace{0.5cm}
\caption{ \label{temp}
The Thompson scattering measurement of temperature, averaged over the
plasma edge region 
(between 3.74m and 3.80m along its line of sight to the magnetic
axis), from all 
ELMs and plasma pulses, synchronised to the ELM times to allow an
estimate for the temperature's pre- and post-ELM evolution to be made. 
Brown circles are individual measurements, the thick black line is
their average, and the dashed black lines are their average $\pm$
the standard deviation. 
Units are eV (vertical axis), and seconds (horizontal
axis). 
Notice that there is a minimum at around 2 milliseconds, as was
similarly found in Refs. \cite{Beurskens,Frassinetti}. 
} 
\end{figure*}

\begin{figure*}[htbp!]
\begin{center}
\includegraphics[width=10cm]{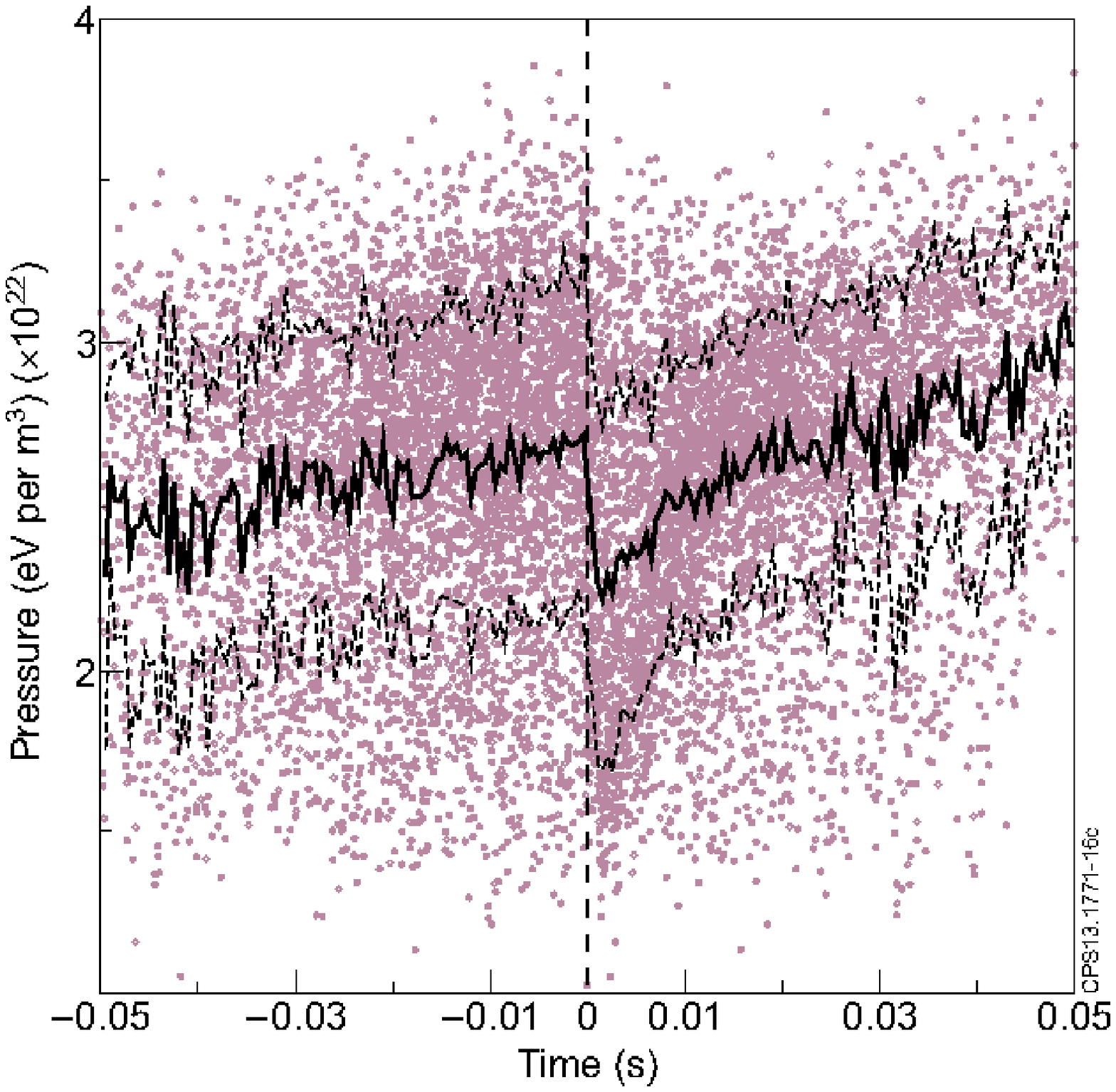}
\end{center}
\vspace{0.5cm}
\caption{ \label{pres}
The Thompson scattering measurement of pressure, averaged over the
plasma edge region 
(between 3.74m and 3.80m along its line of sight to the magnetic
axis), from all 
ELMs and plasma pulses, synchronised to the ELM times to allow an
estimate for the pressure's pre- and post-ELM evolution to be made. 
Brown circles are individual measurements, the thick black line is
their average, and the dashed black lines are their average $\pm$
the standard deviation. 
Units are eV m$^{-3}$ (vertical axis), and seconds (horizontal
axis). 
After 20 milliseconds the pressure is approximately the same as its
pre-ELM value. This partly explains why after 20 milliseconds the ELM
energies are statistically the same. 
} 
\vspace{0.5cm}
\end{figure*}

For the plasmas considered here, the edge plasma properties prevented
JET's ECE diagnostic from providing reliable edge-temperature
measurements. 
Thompson scattering can also provide edge temperature measurements,
but at present only every 50 milliseconds, that compares
with the average time between ELMs of about 30-40 milliseconds for
these pulses. 
This prevents us from determining whether the temperature
and pressure drops after individual
ELMs are dependent on the waiting time since the previous ELM. 
However we can get an approximate estimate for the {\sl average}
changes in edge temperature and pressure before and after ELMs by
synchronising the Thompson scattering data to the ELM times and then
combining all the data into a single plot. 
These plots of temperature, density, and pressure, are in figures
\ref{dens}, \ref{temp}, and \ref{pres}. 
The measurements shown are an average of the Thompson scattering 
measurements between 3.74m and 3.80m along its line of sight to 
the magnetic axis, which 
crosses a similar region to the line-integrated measurements
of figure \ref{PostELMdensity} (that passes vertically downwards 
perpendicularly through the midplane at 3.73m), and 
ignores measurements from the outermost edge where the Thompson
scattering errors are large.  
The density measurements shown in figure \ref{dens} are consistent
with figure \ref{PostELMdensity} (the line integrated density cuts 
through $\sim$1m of plasma, see Appendix \ref{ROG} for more details),
and a minimum in the post-ELM edge density is 
again seen at around 3-5 milliseconds after the ELM. 
The temperature shown in figure \ref{temp} falls to a minimum at 
around 2 milliseconds after an ELM, as was similarly found in 
Refs. \cite{Beurskens,Frassinetti}.   
Following an ELM, figures \ref{dens}-\ref{pres} give the average drop
in density to be of order 
(0.5 $\pm$ 0.5).10$^{19}$m$^{-3}$, the drop in temperature to be of order
(75 $\pm$ 75)eV, and the drop in pressure to be of order 
(0.5 $\pm$ 0.5).10$^{22}$eVm$^{-3}$, where the error bars are the standard
deviation.  
The volume of edge plasma that the measurements cover is of order
15m$^3$, so for a plasma pressure $p$ and volume $V$, the thermal
(``kinetic'') 
energy lost from the region is  $(3/2)\int p dV  \sim$
(3/2)(15).10$^{22}$(1.6).10$^{-19}$J$\sim$36kJ.
The estimate may be a bit smaller than it should be because we have
only considered changes in energy between the flux surfaces that cut
3.74m and 3.8m along the Thompson scattering's line of sight to the
magnetic axis. 
Note however that a 36kJ loss over 2-5 milliseconds is consistent with
the time and magnitude of the first minimum in figure
\ref{PostELMEnergy}, suggesting that EFIT's estimate for the loss in
thermal (``kinetic'') plasma energy is approximately correct during
the first 0-5 milliseconds of an ELM. 
This is reassuring. Any random errors in EFIT's calculation for the
plasma pressure will be eliminated by the subsequent averaging over
large numbers of data sets; the agreement with the Thompson scattering
measurement suggest that any systematic errors over this 0-5
millisecond post-ELM time-period are reasonably small. 
Note that the plasmas here have smaller current, smaller toroidal
magnetic field, smaller heating, and  consequently smaller ELMs than
those in Ref. \cite{Frassinetti}. 
The second minima in figure \ref{PostELMEnergy} at 10 milliseconds 
requires a further drop in energy by 70-100 kJ. 
Because the direct measurement of plasma pressure (figure \ref{pres}),
disagrees with EFIT's calculated plasma pressure during the 5-10
millisecond time period after an ELM, it seems likely that EFIT's
calculations for the pressure during this time period are incorrect. 
As mentioned previously, the cause of the difference between the
direct Thompson scattering measurements and EFIT's calculated pressure
are likely to be due to non-ideal, possibly resistive processes, that
occur while the plasma is relaxing to a new post-ELM equilibrium. 
Returning to figures \ref{dens}, \ref{temp}, and \ref{pres}, it is
clear that after about 20 milliseconds the edge pressure has returned
to very close to its pre-ELM value. 
There continues to be a small increase in
pressure from 20 
milliseconds until the next ELM, but this is small compared with the
scatter in the data. 
This suggests a picture for these pulses where the edge pedestal is
largely restored after 20 milliseconds, which helps to explain why the
ELM energies are statistically similar after 20 milliseconds
(figs. \ref{dE-dt}, \ref{dE-dt-0.005s}, \ref{dn-dt}, \ref{Gauss1}, and
\ref{Gauss2}). 
It also supports a picture where the ELM energy is determined by the
maximum edge pressure. 

\section{Discussion and Conclusions}\label{Concs}

\begin{figure*}[htbp!]
\begin{center}
\includegraphics[width=10cm]{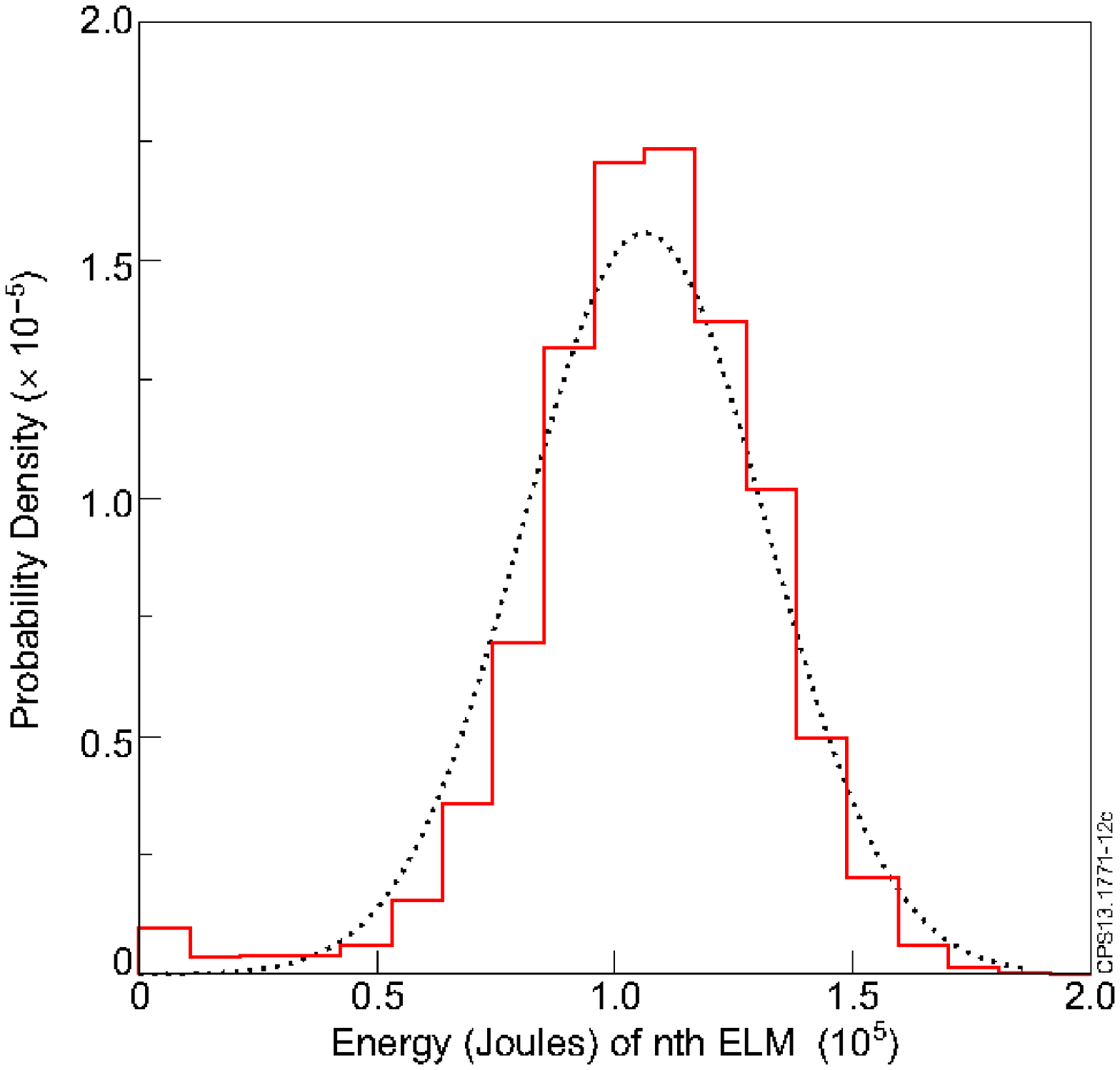}
\end{center}
\caption{ \label{Gauss1}
The probability density of ELM energies calculated with $t_m=0.01$ is
plotted, along with a 
simple Gaussian fit (dotted black curve). 
Even without excluding the ELMs that arise with
waiting times less than roughly 0.02 seconds, the distribution of ELM
energies is approximately Gaussian, with an average ELM energy of
(1.06)10$^5$ 
Joules and a standard deviation of (0.26)10$^5$ Joules, giving a
co-efficient of variation of 0.25 for the spread of ELM energies. 
} 
\end{figure*}

\begin{figure*}[htbp!]
\begin{center}
\includegraphics[width=10cm]{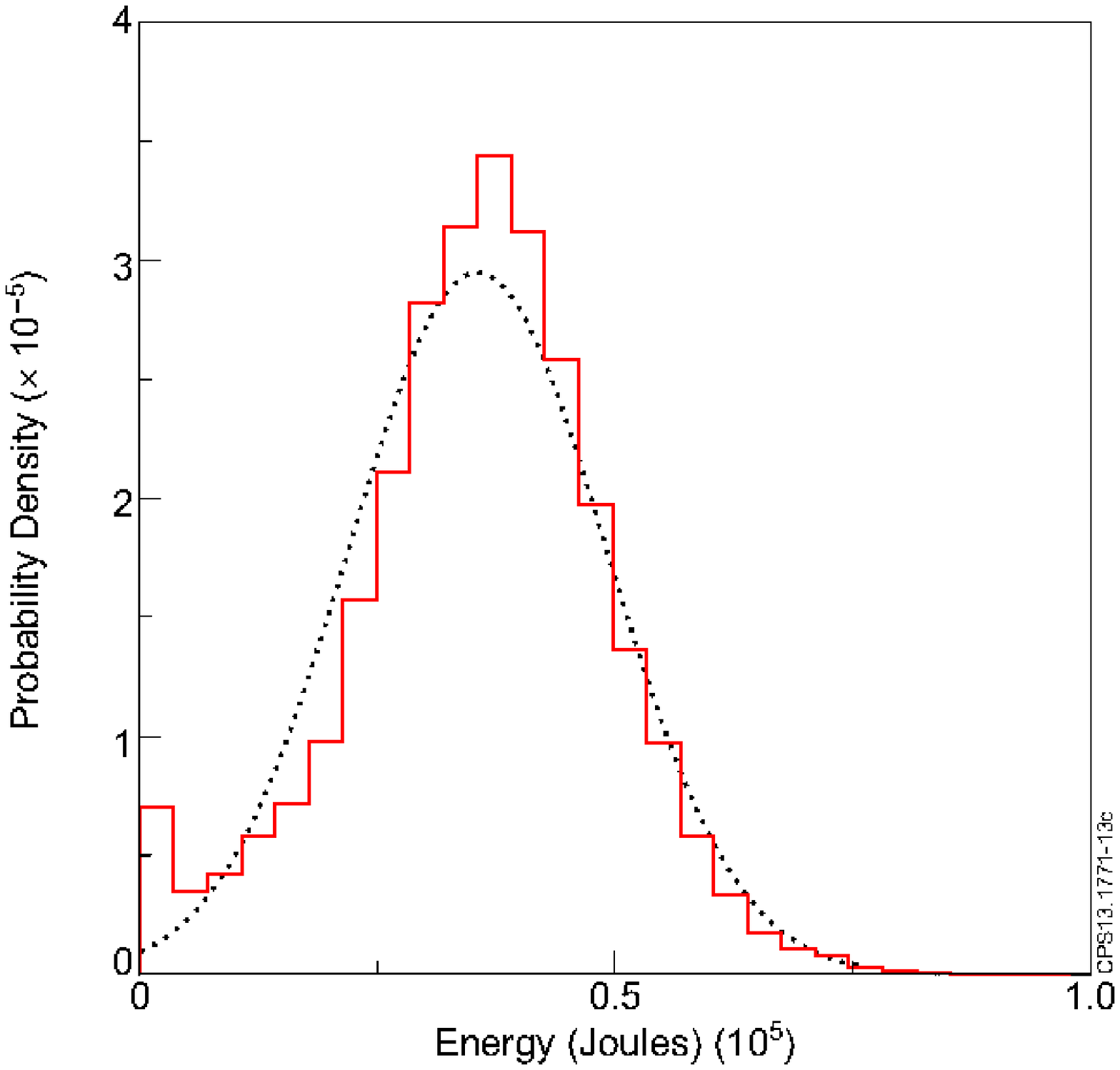}
\end{center}
\caption{ \label{Gauss2}
The probability density of ELM energies calculated with $t_m=0.005$ is
plotted, along with a 
simple Gaussian fit (dotted black curve). 
Even without excluding the ELMs that arise with
waiting times less than roughly 0.02 seconds, the distribution of ELM
energies is approximately Gaussian, with an average ELM energy of
(3.55)10$^4$ 
Joules and a standard deviation of (1.35)10$^4$ Joules, giving a
co-efficient of variation of 0.38 for the spread of ELM energies.  
} 
\end{figure*}

We have used the line integrated edge density and the thermal energy
calculated with EFIT to
study the properties of the $\sim$10,000 ELMs produced from 120 (of
150) almost identical JET pulses, and have used Thompson
scattering to check these results by observing the average evolution
of the edge temperature and pressure in these plasmas. 
It is found that:  
i) There are clear timescales associated with the
ELMs, 
with a loss of edge temperature over 2 milliseconds, a loss of density
and pressure over 5 milliseconds, 
and an additional 10 millisecond timescale over which non-ideal
affects appear to make EFIT's equilibrium reconstruction unreliable.  
The energy losses over the shorter 2-3 milliseconds timescale
appear to be associated with the loss of thermal plasma energy
(``kinetic'' energy), with 
minima in edge temperature, pressure, and density occurring within a
2-5 millisecond timescale after an ELM.
The 0.005-0.01 second timescale is a previously unreported timescale 
during which the (ideal-MHD) plasma pressure reconstructed by EFIT
disagrees with Thompson scattering measurements, 
and is a similar timescale to the 8 milliseconds resistive timescale
of JET's plasma pedestal (see  Appendix \ref{relax}). 
This suggests that after an ELM there are non-ideal, possibly  
resistive processes occurring over a 5-10 millisecond timescale, as the
plasma pressure and edge pedestal 
recover towards their pre-ELM values. 
It also helps to explain why for timescales of order 0-5 milliseconds,
EFIT's calculations and the Thompson scattering measurements agree. 
ii) Following an ELM, no ELMs are
observed until approximately 0.012 seconds later, when they are
statistically about 60\% of the size of ELMs observed in the next
cluster at approximately 0.02 seconds. 
Similar remarks apply regardless of whether the shorter or longer
timescales of $t_m =$ 0.005 seconds or $t_m =$ 0.01 seconds are used
to define the energy drop due to an ELM. 
iii) From 0.02 seconds onwards, the ELM
energies are all statistically similar, with an approximately Gaussian
distribution that is independent of the waiting times between the
ELMs, and a standard deviation that is about 1/4 
of the average ELM energy (see figures \ref{Gauss1} and
\ref{Gauss2}).   
Although the edge pressure appears to increase until an ELM, it
changes very little compared with its rapid recovery in the 20
milliseconds after an ELM. 
This suggests that the edge pedestal is largely recovered 20
milliseconds after an ELM, consistent with the similarities in ELM
energies from 20 milliseconds onwards. 
If the edge pressure and ELM properties are so similar from 20
milliseconds after an ELM, there are some interesting questions about:
what triggers the next ELM? the proximity of the edge plasma to
marginal stability? and whether the ELM trigger is better regarded as
a statistical or deterministic process?

The first point (i), 
helps to clarify the processes taking place during an ELM that need to be
better understood, and includes the observation of an extra relaxation 
time during the ELM process.  
Points (ii)-(iii) have clear consequences for ELM mitigation, at least
for plasmas similar to those discussed here. 
The maximum (natural) ELM frequency that is observed has an ELM waiting
time of approximately 0.012 seconds. 
ELMs with waiting times of $\simeq$ 0.012 seconds have an average 
energy loss associated with the ELM that is roughly 60\% that of the
ELMs with waiting times of 0.02 seconds or longer.  
So presuming that ELM pacing techniques can 
consistently pace ELMs with waiting times of 
0.012 seconds or less, then a reduction in average ELM
energy by at least 40\% seems likely to be possible, or 50\% if we
presume that the
shorter timescale of $t_m =$ 0.005 seconds determines the peak heat
fluxes onto surfaces. 
In principle JET can trigger ELMs with ``vertical kicks'' \cite{NewRef},
with frequencies up to about 100Hz, so it would be possible 
to test this experimentally at JET using 83Hz kicks. 
Although we caution that even at 83Hz, the spread of the ELM energies
observed in figures 6 and 7 can include energies significantly above
the average observed value. 
To the authors' knowledge, no vertical kick experiments have yet been
done at this frequency.  
If a resistive process is responsible for the  
0.01 second timescale, then it might allow the energy to be lost more
uniformly in the form of plasma filaments for example, 
possibly helping to reduce the peak heat fluxes at the divertor. 
The maximum heat fluxes (gradients) in figure \ref{PostELMEnergy} 
are between 0-2 milliseconds and 5-8 milliseconds, 
although only the heat flux calculated over 0-2 milliseconds is
thought to be a reliable estimate. 

The results summarised in figure \ref{dE-dt} clearly fail to satisfy
the often quoted relationship of $\delta E\propto 1/f$.  
This may be partly because the relationship that is measured in such
papers is actually  
$\langle \delta E \rangle \propto 1/ \langle f \rangle$, and
consequently refers to average properties 
of possibly very different plasmas, and not to the properties of
individual ELMs within similar plasmas. 
Unfortunately it is this latter quantity, the relationship between ELM
size and ELM waiting time that is important for ELM mitigation by pacing
techniques.  
Without a reduction in ELM energy, mitigation techniques will need to
reduce either the peak heat flux or increase the wetted area onto
which energy is deposited. 
The results presented here also only represent one particular type of
pulse in one tokamak, JET. 
It is entirely possible that different pulse types or different
machines might have very different ELM statistics. 
The purpose of the analysis here is to provide a robust
analysis of these 2T 2MA pulses for which such large numbers of
(almost) statistically equivalent ELMs
are available, providing a clear indication of ELM behaviour 
for this particular pulse type at least. 
The hope was that the excellent
statistics might indicate new or unexpected ELM physics. 
One of the unexpected results is the observed 
independence of ELM 
size and waiting time for waiting times greater than about 0.02
seconds. 
The generality of these results remains to be
determined, and may require dedicated new experiments to ensure a
robust answer.

The results here have consequences for the correct
construction of models for ELMs and ELMing behaviour. 
For the pulses discussed here, 
beyond the group of ELMs with $\sim$0.012 seconds waiting time, the ELM
waiting times and 
energies are independent. 
Consequently for such ELMs, models to describe their waiting times and
ELM-energy probability distributions can be treated independently. 
Even more surprisingly perhaps, is that figures \ref{Ena} and
\ref{Enb} suggest that the energies of subsequent ELMs are
independent, so that a large ELM is as likely to be followed by
another large ELM as by a small ELM. 
Surprising as this may be, it is likely to make the statistical
modelling of ELM energies considerably easier.  
Clearly, the statistical relationships observed here need to be
reproducable by any 
simulation that is correctly modelling these plasmas. 
Similar remarks apply to the relaxation of the plasma's energy, and
the sequence of processes and timescales by which the plasma loses
energy due to an ELM.

To conclude, we have presented the analysis of an unprecedentedly
large number of statistically equivalent 2T 2MA JET ITER-like wall
H-mode plasmas.  
This has led to the observation of an extra 0.01 second timescale
associated with the ELM process, that is consistent with a resistive
mechanism that allows the plasma to relax to a new post-ELM
equilibrium. 
For the plasmas discussed here, surprising
results are reported about the independence of ELM energy
and frequency, and the independence of energies of consecutive 
ELMs. 
Whether the results found here are more generally true 
is unknown, it may be some time before equivalently large datasets for
different pulse types or from different machines become available.

\begin{acknowledgments}
We would like to thank the referees for  
helping to improve the paper, and to: Joanne
Flanagan for help with Thompson scattering data, 
Martin Valovic for help with resistivity estimates, Ian 
Chapman for comments on the paper, and 
to Howard Wilson and Fernanda Rimini for raising the question 
of how the ELM energies are 
related to the waiting times for this set of pulses. 
The experiments were planned by S. Brezinsek, 
P. Coad, 
J. Likonen, and 
M. Rubel. 
This work, part-funded by the European Communities under the
contract of Association between EURATOM/CCFE was carried out within
the framework of the European Fusion Development Agreement.  
For further information on the contents of this paper please contact
publications-officer@jet.efda.org. 
The views and opinions expressed 
herein do not necessarily reflect those of the European Commission. 
This work was also part-funded by the RCUK Energy Programme [grant
number EP/I501045].
\end{acknowledgments}

\appendix

\section{Plasma motion and measurements}\label{ROG}

Following an ELM there will be a radial motion of the plasma. 
This will modify the measurements in two ways: 
i) the length of plasma that the line-integrated measurements pass
through will reduce slightly, 
ii) the measurements will be of a slightly different region of plasma
due to its small radial displacement. 
Here we will estimate the changes to measurements that would be
expected to result from a small radial displacement of the plasma, and
confirm that they are small compared to the measured changes that
occur after an ELM, and can therefore be 
neglected. 

During an ELM the radial outer gap between the outboard plasma and the
outer wall changes by 7-8mm, which is 0.007-0.008m. 
The line integrated measurement passes through approximately 1.45m of
plasma, approximately 1.1m of which is through the higher density
region above the top of the plasma pedestal (these lengths can be used
to estimate the plasma density from the line-integrated density
measurement).  
Allowing for the geometry of the flux surfaces, a 7-8mm radial shift
will only modify the length of plasma it passes through by a few cms
at the most, or by 1-2\%. 
Therefore because the measured changes in line-integrated density are
of order 10-20\%, we can neglect this effect. 

The edge pedestal is thought to be 2-3cms at most \cite{Frassinetti2},
and the line-integrated measurement cuts through the mid-plane at
about 3.73m, so most of the line of sight is through plasma above the
top of the pedestal (at the mid-plane the plasma edge is at
approximately 3.80m). 
Above the top of the pedestal the plasma density gradient is between
2.10$^{19}$m$^{-4}$ and 5.10$^{19}$m$^{-4}$, so a 0.008-0.009m ROG shift
will change the density that is measured by the line-integrated
measurement by less than
(0.05).10$^{19}$m$^{-3}$, which is less than 1\%. 
Therefore because the measured changes in line-integrated density are
by 10-20\%, we can neglect this affect also. 

In summary, compared with the measured changes in density, the changes
due to the radial plasma shift that occurs with an ELM can be
neglected. 
Similar remarks apply to the temperature and pressure measurements. 

\section{The current relaxation timescale}\label{relax}

As given in Ref. \cite{Friedberg} for example, the plasma's
resistivity is, 
\begin{equation}\label{eta}
\eta = (6.5) 10^{-8} \left( \frac{1}{T_k^{3/2}} \right) \mbox{  $\Omega$ m}
\end{equation}
where $T_k$ is the plasma's temperature measured in electron Volts. 
A multiplicative constant modifies Eq. \ref{eta} when Neoclassical
effects are included and if $Z_{eff}\neq 1$, but Eq. \ref{eta} is a
reasonable order of magnitude estimate. 
The resistive MHD equations \cite{Friedberg} give, 
\begin{equation}
\frac{\partial \vec{B}}{\partial t} = 
\left( \frac{\eta}{\mu_0} \right) \nabla^2 \vec{B} 
\end{equation}
from which a dimensional analysis gives the resistive timescale $\tau$
as, 
\begin{equation}\label{tau}
\tau \sim \left( \frac{\mu_0}{\eta} \right) L^2 
\end{equation}
where $L$ is  a typical length scale and $\mu_0 = (4\pi) 10^{-7}$ Farad
m$^{-1}$. 
Combining equations \ref{eta} and \ref{tau} gives, 
\begin{equation}\label{tau2}
\tau \sim (6.2)\pi T_k^{3/2} L^2 
\end{equation}
Substituting the pedestal width \cite{Frassinetti2} of $L \sim 0.03$m
and temperature at the pedestal's top of $T_k \sim 0.6$keV, gives
$\tau \sim 8$ milliseconds, very similar to the $10$ millisecond
timescale observed in figures \ref{PostELMEnergy} and \ref{dE-tm}.

\end{document}